\documentclass[aps,prb,a4paper,10pt,twocolumn,showpacs,floatfix,superscriptaddress,preprintnumbers,longbibliography]{revtex4-2}
\usepackage{float}
\usepackage{dcolumn,graphicx,color,booktabs,microtype,afterpage}
\usepackage{amssymb}
\usepackage{amsmath}
\usepackage[charter,greekuppercase=italicized]{mathdesign}
\usepackage{sidecap}
\usepackage[mathlines]{lineno}
\usepackage{gensymb}

\graphicspath{{./}{figure/}}
\renewcommand{\tablename}{Table}
\makeatletter\renewcommand{\fnum@figure}[1]{\figurename~\thefigure.~}\makeatother
\makeatletter\renewcommand{\fnum@table}[1]{\tablename~\thetable.}\makeatother

\newcount\hh \newcount\mm
\hh=\time \divide\hh by 60
\mm=\hh \multiply\mm by 60 \mm=-\mm
\advance\mm by \time
\def\now{\number\hh:\ifnum\mm<10{}0\fi\number\mm}

\usepackage[colorlinks,plainpages=false,linkcolor=blue,urlcolor=blue,citecolor=blue,pdfpagemode=UseNone,pdfstartview=FitBH]{hyperref}
\usepackage{nicefrac}

\newcommand{\tcr}[1]{\textcolor{black}{#1}}

\begin{document}
	\makeatletter\renewcommand{\ps@plain}{%
		\def\@evenhead{\hfill\itshape\rightmark}%
		\def\@oddhead{\itshape\leftmark\hfill}%
		\renewcommand{\@evenfoot}{\hfill\small{--~\thepage~--}\hfill}%
		\renewcommand{\@oddfoot}{\hfill\small{--~\thepage~--}\hfill}%
	}\makeatother\pagestyle{plain}
	
	\preprint{\textit{Preprint: \today, \now}} 
 
\title{Absence of topological Hall effect in Fe$_x$Rh$_{100-x}$ epitaxial films:\\ revisiting their phase diagram}

\author{Xiaoyan Zhu}
\affiliation{Key Laboratory of Polar Materials and Devices (MOE), School of 
Physics and Electronic Science, East China Normal University, Shanghai 200241, China}

\author{Hui Li}
\affiliation{Cavendish Laboratory, University of Cambridge, J J Thomson Avenue, 
	Cambridge, CB3 0HE, UK}

\author{Jing Meng}
\affiliation{Key Laboratory of Polar Materials and Devices (MOE), School of 
	Physics and Electronic Science, East China Normal University, Shanghai 
	200241, 
	China}

\author{Xinwei Feng}
\affiliation{Key Laboratory of Polar Materials and Devices (MOE), School of 
	Physics and Electronic Science, East China Normal University, Shanghai 
	200241, 
	China}

\author{Zhixuan Zhen}
\affiliation{Key Laboratory of Polar Materials and Devices (MOE), School of 
	Physics and Electronic Science, East China Normal University, Shanghai 
	200241, 
	China}
\author{Haoyu Lin}
\affiliation{Key Laboratory of Polar Materials and Devices (MOE), School of 
	Physics and Electronic Science, East China Normal University, Shanghai 
	200241, 
	China}

\author{Bocheng Yu}
\affiliation{Key Laboratory of Polar Materials and Devices (MOE), School of 
	Physics and Electronic Science, East China Normal University, Shanghai 
	200241, 
	China}

\author{Wenjuan Cheng}
\affiliation{Key Laboratory of Polar Materials and Devices (MOE), School of 
	Physics and Electronic Science, East China Normal University, Shanghai 
	200241, 
	China}

\author{Dongmei Jiang}
\affiliation{Key Laboratory of Polar Materials and Devices (MOE), School of %
	Physics and Electronic Science, East China Normal University, Shanghai 
	200241, 
	China}

\author{Yang Xu}
\affiliation{Key Laboratory of Polar Materials and Devices (MOE), School of 
	Physics and Electronic Science, East China Normal University, Shanghai 
	200241, China}

\author{Tian Shang \thanks{Corresponding author: tshang@phy.ecnu.edu.cn}}
\email{tshang@phy.ecnu.edu.cn}
\affiliation{Key Laboratory of Polar Materials and Devices (MOE), School of 
Physics and Electronic Science, East China Normal University, Shanghai 200241, China}
\affiliation{Chongqing Key Laboratory of Precision Optics, Chongqing Institute of East China Normal University, Chongqing 401120, China}

\author{Qingfeng Zhan \thanks{Corresponding author: qfzhan@phy.ecnu.edu.cn}}
\email{qfzhan@phy.ecnu.edu.cn}
\affiliation{Key Laboratory of Polar Materials and Devices (MOE), School of 
Physics and Electronic Science, East China Normal University, Shanghai 200241, 
China}

\begin{abstract}

A series of Fe$_x$Rh$_{100-x}$ ($30 \leq x \leq 57$) films were epitaxially grown using magnetron sputtering, and were 
systematically studied by 
magnetization-, electrical resistivity-, and Hall resistivity measurements.	
After optimizing the growth conditions, phase-pure Fe$_{x}$Rh$_{100-x}$ films were obtained, and their magnetic phase diagram was revisited. 
The ferromagnetic (FM) to antiferromagnetic (AFM) transition is limited at narrow Fe-contents with $48 \leq x \leq 54$ in the 
bulk Fe$_x$Rh$_{100-x}$ alloys.
By contrast, 
the FM-AFM transition in the Fe$_x$Rh$_{100-x}$ films is extended to cover a much wider $x$ range between 33\% and 53\%, 
whose critical temperature slightly decreases as increasing the Fe-content.
The resistivity jump and magnetization drop at the FM-AFM transition are much more significant in the Fe$_x$Rh$_{100-x}$ films with $\sim$50\% Fe-content than in the Fe-deficient films, 
the latter have a large amount of paramagnetic phase. The magnetoresistivity (MR)
is rather weak and positive in the AFM state, while it becomes negative when the FM phase 
shows up, and a giant MR appears in the mixed FM- and AFM states. 
The Hall resistivity is dominated by the ordinary Hall effect in the AFM state, while in the mixed state or high-temperature 
FM state, the anomalous Hall effect takes over. The absence of topological Hall resistivity in Fe$_{x}$Rh$_{100-x}$ films 
with various Fe-contents implies that the previously observed topological Hall effect is most likely extrinsic. We 
propose that the anomalous Hall effect caused by the FM iron moments at the interfaces nicely explains the hump-like anomaly 
in the Hall resistivity. Our systematic investigations may offer valuable insights into the spintronics based on iron-rhodium 
alloys.  

\end{abstract}

\maketitle

\section{INTRODUCTION}

The CsCl-ordered equiatomic iron-rhodium (Fe-Rh) alloy undergoes a first-order magnetic phase transition from the 
high-temperature ferromagnetic (FM) 
state to the low-temperature antiferromagentic (AFM) state near room temperature~\cite{Uhl_2016,Li_2022}.  
Such a transition leads to a significant drop in the magnetization and a jump in the 
electrical resistivity, which can be applied to the spintronic 
devices. The FM-AFM transition in Fe-Rh alloys can be easily tuned by external control 
parameters, such as chemical 
substitution~\cite{Kinane_2014}, epitaxial 
strain~\cite{Arregi_2020,Xie_2017,QIAO_2019}, and magnetic or electric 
fields~\cite{Cherifi_2014,Maat_2005,Lee_2015,Liu_2016}. 
Many exotic properties have been found in Fe-Rh alloys, which are closely related to their FM-AFM transition.
The spin–orbit torque 
efficiency can be significantly tuned by varying the temperature across the FM-AFM transition in Fe-Rh-based heterostructures~\cite{Cao_2022}.
The large magnetocaloric effect can be controlled by ferroelectric 
domains in Fe-Rh film near the FM-AFM transition~\cite{QIAO_2020}. 
Since the FM-AFM transition presents near the room temperature, therefore, Fe-Rh 
alloys represent one of the ideal candidate materials for spintronic 
applications, such as memory resistor~\cite{Marti_2014}, heat-assisted magnetic 
recording~\cite{Thiele_2004}, and magnetic refrigeration~\cite{QIAO_2020}.   

\begin{figure}[!htp]
	\centering
	\includegraphics[width = 0.48\textwidth]{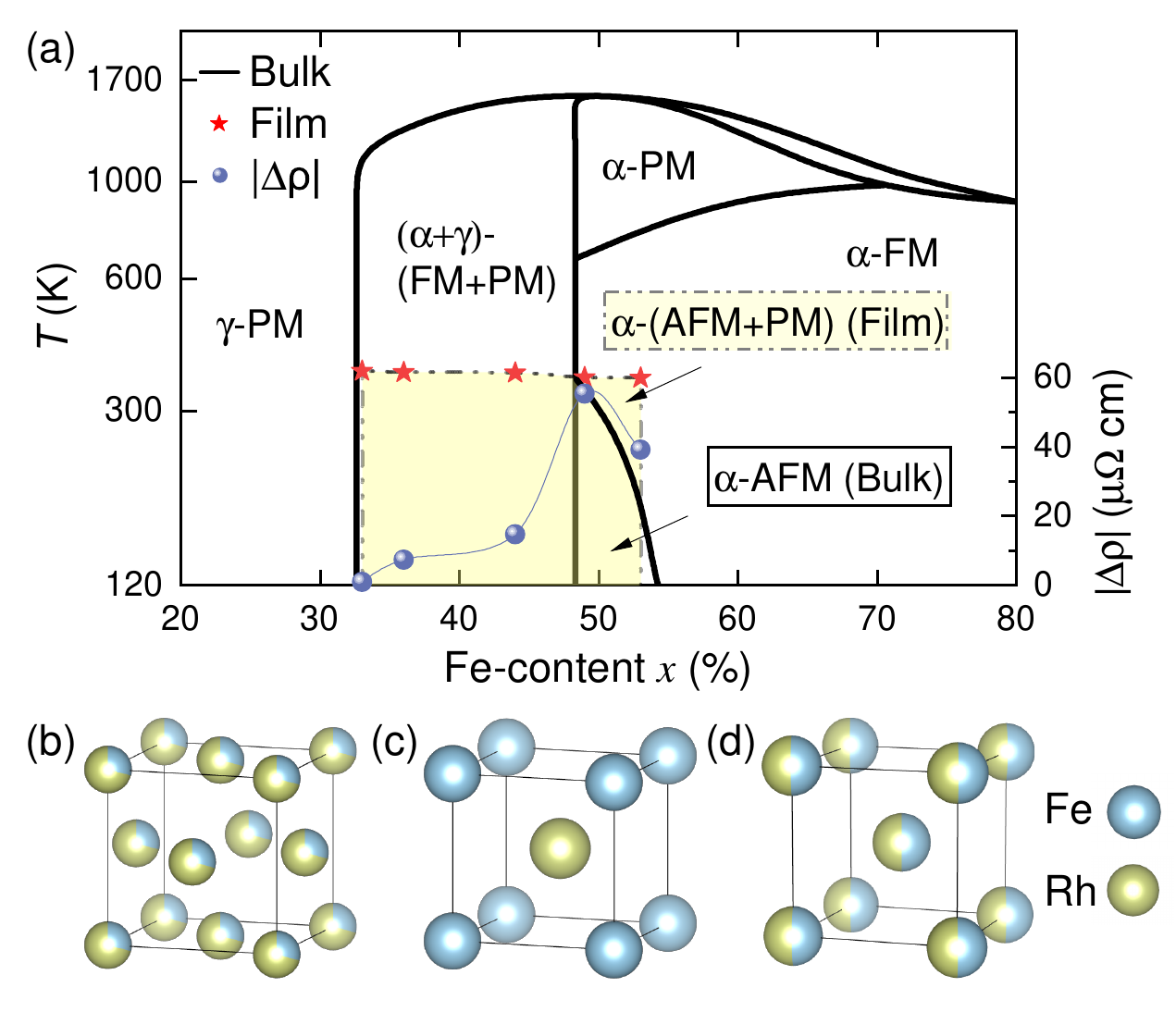}
	\vspace{-3ex}%
	\caption{\label{fig:fig1}(a) Phase diagram of bulk Fe-Rh alloys and epitaxial Fe-Rh films. Data of bulk alloys were taken 
	from 
	Refs.~\onlinecite{Shirane1963,1982Iron}. 
			The star symbols represent the FM-AFM magnetic transition 
			temperatures (left axis) for Fe-Rh films, while the sphere symbols 
			show the resistivity jump  $\lvert$$\Delta$$\rho$$\rvert$ (right 
			axis) for these films against the Fe-content. 
			 Crystal structures for (b) $\gamma$-phase (FCC, $Fm$-3$m$, No.~225) and (c) $\alpha$-phase (BCC, 
			$Pm$-3$m$, No.~221) Fe-Rh. The crystal structure of disordered 
			$\alpha$-phase is shown in panel (d). 
			Different from the ordered $\alpha$-phase in (c), the Fe or Rh 
			atoms occupy the same site in the disordered 
			$\alpha$-phase and $\gamma$-phase.}
\end{figure}

The bulk Fe-Rh alloys exhibit a rich phase diagram when varying the Fe- or Rh concentrations. 
We summarize the phase diagram of bulk Fe-Rh alloys in Fig.~\ref{fig:fig1}(a).
On the Rh-rich side, the $\gamma$-PM indicates the paramagnetic (PM) phase with a face-centered cubic (FCC) crystal structure, where both Rh- and Fe atoms occupy the same sites [see Fig.~\ref{fig:fig1}(b)].
For the intermediate Fe concentration ($<$ 48\%), the Fe-Rh alloys adopt the mixed $\alpha$- and $\gamma$-phases. While the $\gamma$-phase remains PM, the $\alpha$-phase becomes FM 
below certain temperatures [denoted as ($\alpha$+$\gamma$)-(FM+PM) in Fig.~\ref{fig:fig1}(a)]. When increasing the Fe-content above 48\%, the Fe-Rh alloys show a pure $\alpha$-phase with a body-centered cubic (BCC) crystal structure. In particular, 
the Fe-Rh alloys with 48-54\% Fe-content undergo multiple magnetic transitions, from high-temperature PM state (marked as $\alpha$-PM) to the FM state ($\alpha$-FM), and then finally to the low-temperature AFM state ($\alpha$-AFM). The FM-AFM transition temperature decreases as increasing the Fe-content.  
For these alloys, the Fe and Rh atoms occupy the corner- and center sites, 
respectively [see Fig.~\ref{fig:fig1}(c)]. It is noted that the  $\alpha$-FM and $\alpha$-AFM are also known as ordered 
$\alpha$'-phase and  $\alpha$''-phase. 
On the Fe-rich 
side, both Rh- and Fe atoms 
occupy the same sites [see Fig.~\ref{fig:fig1}(d)],  
and the Fe-Rh alloys behave similarly to pure Fe metal, 
exhibiting a FM ground state below the Curie temperature 
($\sim$1000\,K).
In the $\alpha$-AFM phase, the Fe-moments are aligned with a collinear G-type magnetic structure, exhibiting a typical magnetization value of $\sim$3.1\,$\mu_\mathrm{B}$~\cite{Bertaut1962}. However, there is no net moment on the Rh site. While in the $\alpha$-FM phase, both Fe-moments ($\sim$3.2\,$\mu_\mathrm{B}$) and Rh-moments ($\sim$0.9\,$\mu_\mathrm{B}$) are aligned 
ferromagnetically along the <001>-direction~\cite{Shirane1963}. 

In addition to the exotic properties related to the FM-AFM transition in the
Fe-Rh alloys~\cite{Zhu_2022JPCM,ZHU_2022JAC,Liu_large_2016}, 
the topological Hall effect (THE) has been observed recently in equiatomic Fe-Rh thin films~\cite{APL_THE}.
In general, the THE is often considered as the hallmark of spin textures with a finite scalar spin 
chirality, e.g., magnetic 
skyrmions~\cite{Matsuno_interface-driven_2016,Neubauer_2009}.
The THE in Fe-Rh film is proposed to be attributed to 
the emergence of noncollinear spin texture arising from the 
competition among different exchange interactions in its AFM state.  
The interfacial inhomogeneity in the magnetic thin film 
could lead to an inhomogeneous anomalous Hall 
effect (AHE), whose 
signal resembles the 
THE~\cite{FAKE_THE,Fake_THE2,FAKE_THE3,Zhang_2022}. 
Considering that a large amount of the remaining FM phase persists in the AFM state of Fe-Rh film in the previous work~\cite{APL_THE},
the origin of THE requires further investigation.  
In addition, while most of the thin-film studies focus on equiatomic Fe-Rh films (i.e., $\sim$50\% Fe-content), less is known for Fe-Rh films with different Fe-contents.   

Here, we revisit the phase diagram of Fe-Rh epitaxial thin films 
by varying the Fe- or Rh-contents, and report
a comprehensive study of their magnetic and transport properties by means of magnetization-, electrical resistivity-, and 
Hall resistivity measurements.
Different from the bulk alloys, the Fe-Rh films show a pure $\alpha$-phase in a 
wide Fe-concentration range (i.e., 33 to 53\%).   
The absence of topological Hall resistivity in our high-quality epitaxial Fe-Rh films with different Fe-contents excludes its possible nontrivial origin. 
We propose that the anomalous Hall resistivity caused by the remaining FM moments could give rise to a THE-like signal in the 
Hall resistivity.

\section{EXPERIMENTAL DETAILS}

A series of Fe$_x$Rh$_{100-x}$ ($30\leq x \leq57$) films with a thickness of $\sim$50\,nm were epitaxially grown on 
(001)-oriented  MgO substrates by magnetron co-sputtering Fe and Rh targets in an 
ultrahigh vacuum chamber with a base pressure lower than 1$\times$10$^{-8}$ Torr.
To remove surface contamination, MgO substrates were pre-annealed at 600\,$^\circ$C for 1\,h in 
the vacuum. Afterwards, the substrates were heated up to 700\,$^\circ$C, where 
both Fe and Rh atoms were deposited under a 3\,mTorr-Ar pressure. 
During the deposition, MgO substrates were continuously rotated to improve homogeneity. 
After the deposition, Fe$_x$Rh$_{100-x}$ films were annealed in situ at 750\,$^\circ$C for an extra hour 
to improve their crystallinity. 
Finally, a 3-nm-thick Ta cap layer was deposited at room temperature to avoid oxidation.

The crystal structure and the epitaxial nature of Fe$_x$Rh$_{100-x}$ films were characterized by Bruker D8 Discover high-resolution x-ray diffractometer (HRXRD).
The thickness of films were determined by x-ray reflectivity (XRR). 
The measurements of electrical resistivity ($\rho$), Hall resistivity ($\rho_\mathrm{xy}$), and 
magnetization were performed on a Quantum Design physical property measurement 
system (PPMS) and a  magnetic property measurement system (MPMS), respectively.   
For the transport measurements, the Fe$_x$Rh$_{100-x}$ films were patterned 
into a Hall-bar geometry (central area: 
0.2\,mm $\times$ 4\,mm; electrodes: 0.4\,mm $\times$ 0.65\,mm) by using a 
shadow mask during the growth. 
To avoid spurious resistivity contributions due to misaligned
Hall probes, the longitudinal contribution to the Hall resistivity
$\rho_\mathrm{xy}$, was removed by an anti-symmetrization procedure,
i.e., $\rho_\mathrm{xy}(H)$ = [$\rho_\mathrm{xy}(H)$ -- $\rho_\mathrm{xy}$(--$H$)]/2. Similarly, in the case of
longitudinal electrical resistivity $\rho$ measurements, the spurious
transverse contribution was removed by a symmetrization
procedure, i.e., $\rho(H)$ = [$\rho(H)$ + $\rho$(--$H$)]/2.

\section{RESULTS AND DISCUSSIONS}

\subsection{x-ray diffraction and lattice parameters}

We estimated the composition of Fe$_x$Rh$_{100-x}$ films using the following model:
\begin{equation}
 \label{eq:thickness}
 x = n\cdot N_{A} = (m/M_\mathrm{A}) \cdot N_\mathrm{A} =
[(\beta \cdot v \cdot t \cdot S) /M_\mathrm{A}]  \cdot N_\mathrm{A}.
\end{equation}
Here, $n$, $m$, $N_{A}$, and $M_{A}$ are molar number, mass, Avogadro constant, and molar mass;
$v$ and $t$ are deposition rate and time;
$\beta$ and $S$ represent the density and surface area of the film, respectively.
The deposition rate $v$ was controlled by adjusting the DC sputtering power of Fe and Rh targets, which was further calibrated by XRR 
measurements.
Table~\ref{tab:parameter} lists the sputtering power of Fe- and Rh targets for different Fe$_x$Rh$_{100-x}$ films. 
For instance, to produce the Fe$_{30}$Rh$_{70}$ film, the $P_\mathrm{Fe}$ and $P_\mathrm{Rh}$ were set to 20 and 15\,W, respectively. 
We compared the magnetization- and electrical-resistivity results of the Fe$_{49}$Rh$_{51}$ film prepared by co-sputtering method with the one grown from Fe$_{50}$Rh$_{50}$ alloy target, 
both films show almost identical behaviors, suggesting that the above model gives correct Fe and Rh contents.
Seven Fe$_x$Rh$_{100-x}$ films with $x$ ranging from 30 to 57 were deposited. 

\begin{table}[!th]
	\centering
	\caption{~Summary of the sputtering power of Fe ($P_\mathrm{Fe}$) and Rh ($P_\mathrm{Rh}$) targets  for Fe$_x$Rh$_{100-x}$ thin-film growth and the estimated Fe-content for the obtained films.
		The deviation of Fe-content is about 2\%. Except Fe$_{30}$Rh$_{70}$ film, all other Fe-Rh films adopt a pure $\alpha$-phase. 
		\label{tab:parameter}} 
	\begin{ruledtabular}
		\begin{tabular}{lccccccccc}
			$P_\mathrm{Fe}$ (W) & 20 & 25 & 30 & 35 & 40 & 45 & 35      \\ [1mm]
			$P_\mathrm{Rh}$ (W) & 15 & 15 & 15 & 15 & 15 & 15 & 10      \\ [1mm] 
			$x$ (Fe-content) & 30 & 33 & 36 & 44 & 49 & 53 & 57         \\ 
		\end{tabular}	
	\end{ruledtabular}
\end{table}

\begin{figure}[!htp]
	\centering
	\includegraphics[width=0.48\textwidth]{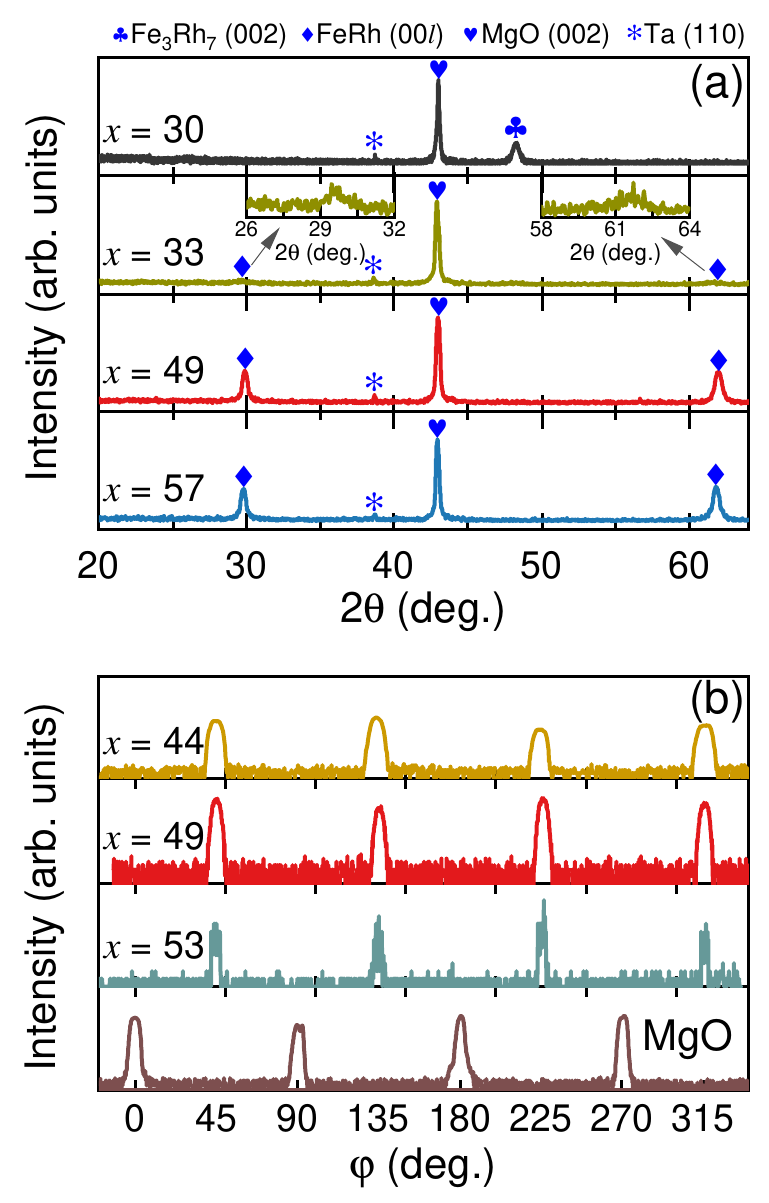}
	\vspace{-6ex}%
	\caption{\label{fig:fig2} (a) Representative XRD patterns of Fe$_x$Rh$_{100-x}$ films for $x$ = 30, 33, 49, and 57.
			The insets show enlarged plots of (001)- and (002)-reflections of Fe$_\mathrm{33}$Rh$_\mathrm{67}$ film. (b) $\varphi$-scan measurements for some selected Fe$_x$Rh$_{100-x}$ films. The intensity is plot on the logarithmic scale.}
\end{figure}

The HRXRD measurements were performed to check the crystal structure and the epitaxial nature of the deposited 
Fe$_x$Rh$_{100-x}$ films.
Figure~\ref{fig:fig2}(a) shows representative XRD patterns for Fe$_x$Rh$_{100-x}$ films with $x$ = 30, 33, 49, and 57.
For $x$ = 30, the (002)-reflection of Fe$_3$Rh$_7$-phase is clear, which adopts a 
$\gamma$-phase~\cite{Inoue_2008} and is 
consistent with the bulk phase diagram in Fig.~\ref{fig:fig1}. 
Since the $\gamma$-phase Fe$_{30}$Rh$_{70}$ is paramagnetic, its 
magnetic- and electrical transport properties will not be discussed here.
No sign of the $\alpha$-phase can be identified in this film. When increasing the 
Fe-content up to 33\%, the $\gamma$-phase 
disappears, in the meanwhile, $\alpha$-phase starts to show up [see insets in Fig.~\ref{fig:fig2}(a)]. 
For $33 \le x  \le 57$, all Fe$_x$Rh$_{100-x}$ films show a 
pure $\alpha$-phase, exhibiting distinct (001)- and (002) reflections. 
This is obviously different from the bulk materials. In the bulk form, 
Fe-Rh alloys (with x $<$ 48) show mixed 
$\gamma$- and $\alpha$-phases.
The absence of foreign phases or misorientation suggests the good quality of our deposited Fe$_x$Rh$_{100-x}$ films.
It is noted that the $\alpha$-phase Fe$_x$Rh$_{100-x}$ films were
epitaxially grown on the MgO substrates with an in-plane 
45$^\circ$~rotation, i.e., FeRh[110](001)||MgO[100](001)~\cite{Xie_2017}, which was further checked by $\varphi$-scan measurements [see Fig.~\ref{fig:fig2}(b)].
It is noted that, for $x$ = 33, the intensities of the XRD reflections are rather low due to the increased mismatch between the film and the substrate, its epitaxial nature cannot be verified by the $\varphi$-scan measurements. 
For $x$ = 33, the epitaxy is less good than the rest of the films, and it might be polycrystalline in nature but with preferred (00$l$) orientation.

\begin{figure}[!htp]
	\centering
	\includegraphics[width=0.48\textwidth]{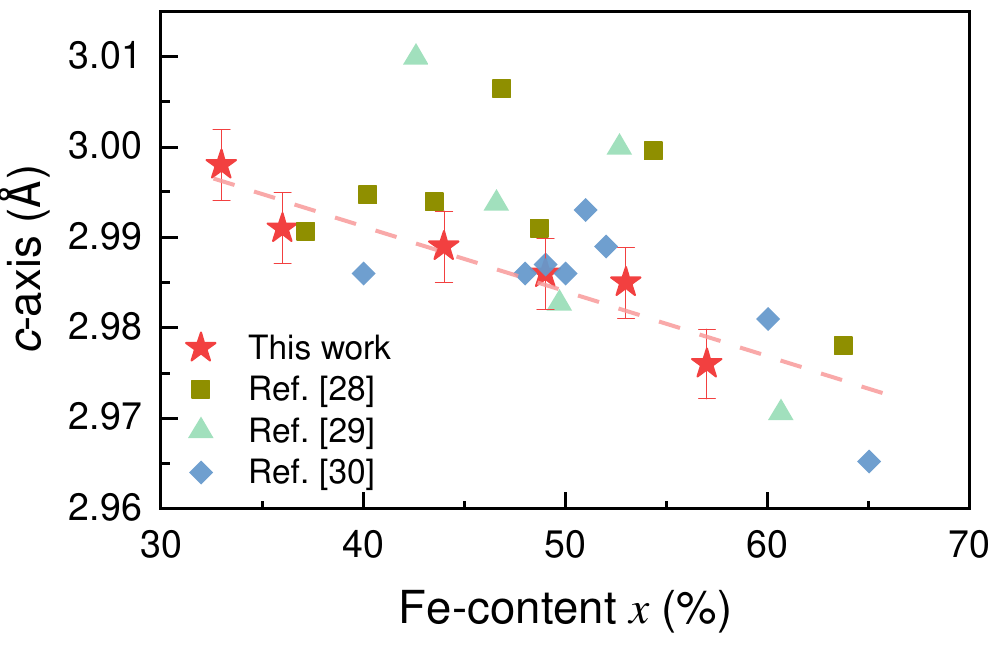}
	\vspace{-6ex}%
	\caption{\label{fig:fig3} Out-of-plane lattice parameters for the Fe$_x$Rh$_{100-x}$ films and bulk counterparts as 
			a function of the 
			Fe-content. 
			The star symbols represent the current work, while the other symbols stand for the previous studies, which were 
			taken from Refs.~\onlinecite{Inoue_2008,Mei_2018,1963Mssbauer}. 
			The dashed line is a guide to the eyes.}
\end{figure}

We estimated the out-of-plane lattice constant (i.e., $c$-axis) for the 
$\alpha$-phase Fe$_x$Rh$_{100-x}$ ($33 \le x \le 
57$) films according to the XRD patterns.
As shown in Fig.~\ref{fig:fig3}, for $x$ = 49, the lattice 
parameter (2.989~\AA) is almost identical to the value of  
Fe$_{50}$Rh$_{50}$ film (2.988~\AA) grown by using a Fe$_{50}$Rh$_{50}$ alloy target~\cite{xie_2020},
which further proves that the above model [see Eq.~\eqref{eq:thickness}] 
estimates 
the proper Fe- or Rh concentration. The obtained lattice parameter 
linearly decreases as increases the Fe-content $x$ (see star 
symbols). 
While in the previous studies, the lattice parameters are clearly more scattered (see square and triangle symbols).
Such linear $x$-dependent lattice parameters again confirm that our Fe$_x$Rh$_{100-x}$ films are very homogeneous and have a 
better quality. 

\subsection{Magnetic properties of Fe$_x$Rh$_{100-x}$ films}
%

%
\begin{figure}[!htp]
	\centering
	\includegraphics[width = 0.45\textwidth]{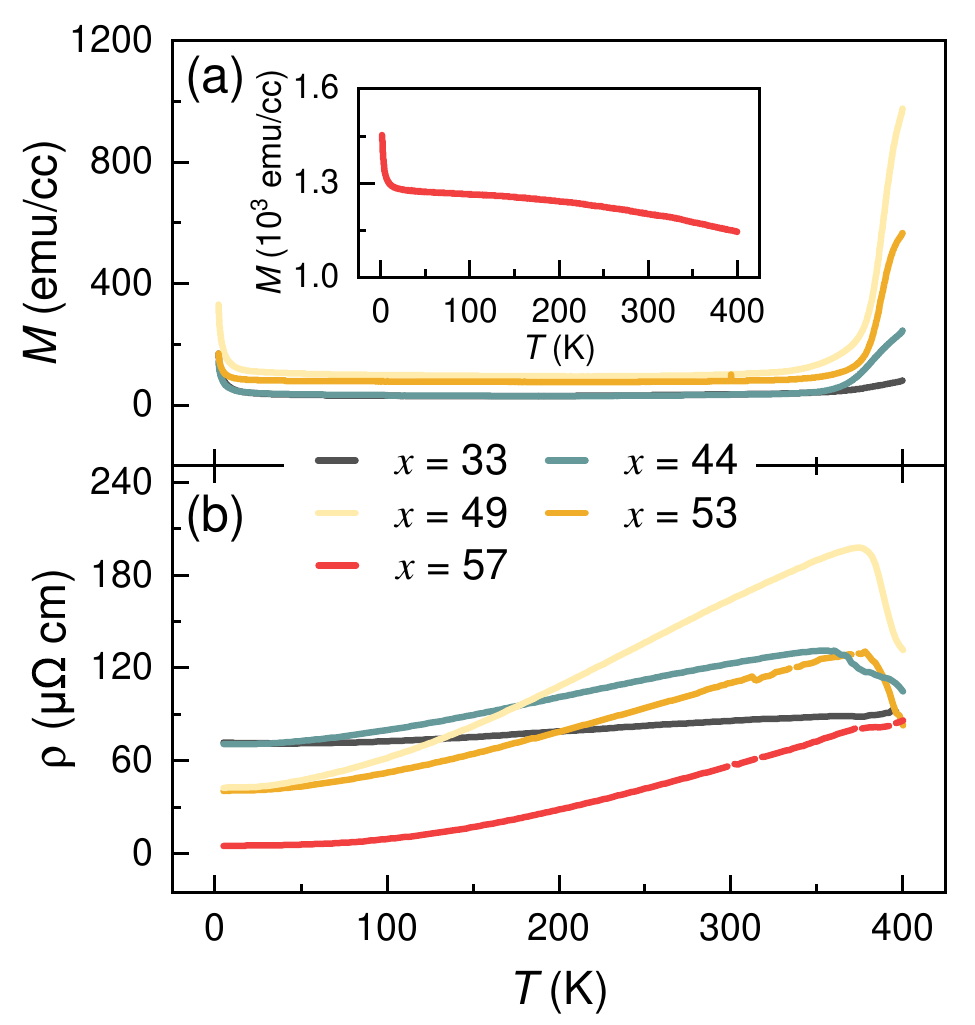}
	\vspace{-3ex}%
	\caption{\label{fig:fig4} Temperature-dependent magnetization 
			collected in a field of $\mu_0H$ = 0.1\,T (a) and zero-field electrical resistivity (b)  
			for Fe$_x$Rh$_{100-x}$ ($33 \le x \le 57$) films. The inset shows 
			the magnetization data for $x$ = 57 film.}
\end{figure}

\begin{figure*}[!htp]
	\centering
	\includegraphics[width = 0.95\textwidth]{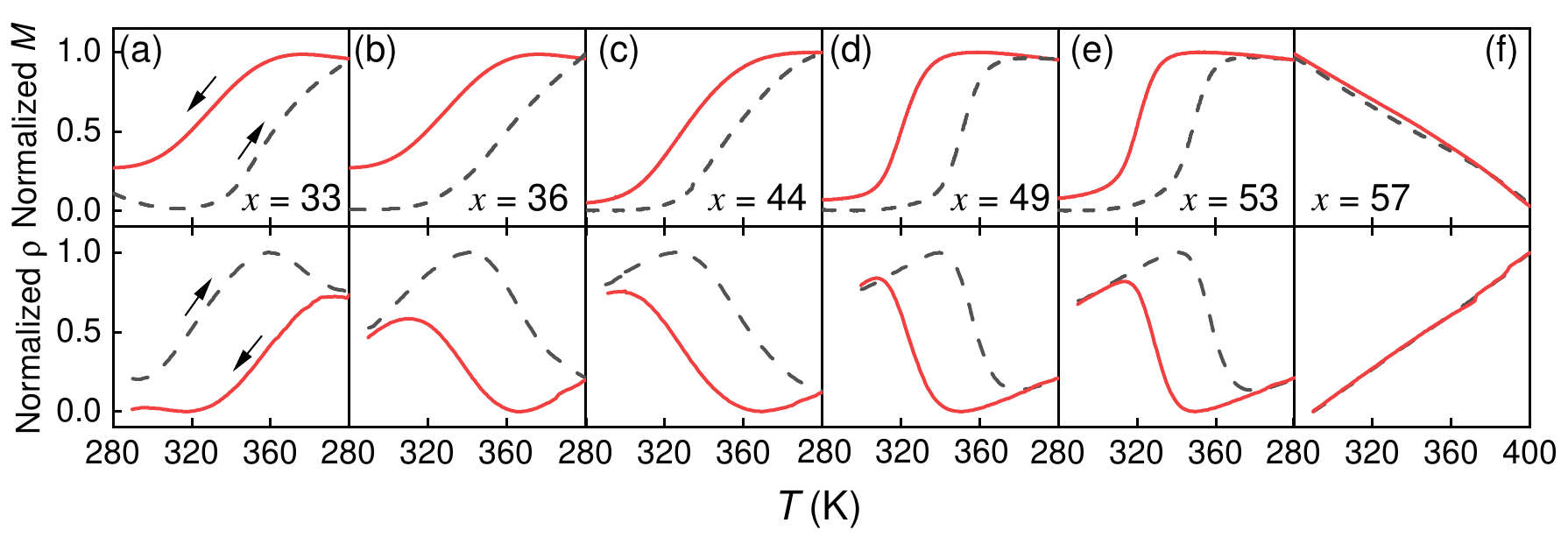}
	\vspace{-3ex}
	\caption{\label{fig:fig5} \tcr{Temperature dependence of the magnetization 
	(up-panels) and the electrical resistivity (bottom-panels) for 
	Fe$_x$Rh$_{100-x}$ ($33 \leq x \leq 57$) films. 
		All the data were collected in a magnetic field of
		$\mu_0H$ = 5\,T during the cooling (solid lines) and heating processes 
		(dashed lines).	To better 
		\tcr{compare the  
		results of different films}, both the magnetization and electrical resistivity data are 
		normalized to the values between 0 and 1.} 
		}  
\end{figure*}

The Fe$_x$Rh$_{100-x}$ ($33 \le x \le 57$) films were first characterized by temperature-dependent magnetization $M(T)$ and 
electrical resistivity $\rho(T)$. 
For the $\alpha$-phase Fe$_x$Rh$_{100-x}$ ($33 \le x \le 53$) films, 
there is a clear anomaly around 380\,K, which is attributed
to the FM-AFM transition. Since the onset of this transition is above 400\,K in zero-field condition in Fe$_x$Rh$_{100-x}$ films, 
the entire transition can not be detected up to 400\,K. However, 
the magnetic field can efficiently tune such a FM-AFM transition, and thus,
the full transition can be clearly seen in a field of 5\,T. 
For $x$ = 57, the bulk sample undergoes a FM 
transition at very high temperature ($\sim$1000\,K) (see Fig.~\ref{fig:fig1}). 
As shown in the inset of Fig.~\ref{fig:fig4}(a),
the magnetization of Fe$_{57}$Rh$_{43}$ film resembles the pure Fe 
film~\cite{Bergqvist_2018}, and there is no magnetic transition below 400\,K.
The magnetization of Fe$_{57}$Rh$_{43}$ film ($\sim$1200\,emu/cc) is almost 
10 times larger than the Fe$_x$Rh$_{100-x}$ ($33 \le x \le 53$) films in 
their AFM state ($\sim$100\,emu/cc). 
It is noted that in all the Fe$_x$Rh$_{100-x}$ 
films, the upturn feature below 10\,K is most likely attributed to the 
PM contribution of MgO substrate~\cite{cabassi_differential_2010}.

Figure \ref{fig:fig4}(b) presents the zero-field temperature-dependent 
electrical resistivity  $\rho(T)$ for Fe$_x$Rh$_{100-x}$  ($33 \le x \le 57$) 
films. 
All the films show a  typical metallic behavior below 350\,K, the electrical resistivity decreases as lowering the temperature. 
Similar to the magnetization results, the resistivity jump at FM-AFM transition is not completed up to 400\,K for $x$ = 44, 
49, 53. For $x$ = 33, though the resistivity anomaly is very weak, 
it is still can be observed (see Fig.5). While for $x$ = 57, 
there is no clear anomaly in the studied temperature range, consisting with its magnetization data [see inset in 
Fig.~\ref{fig:fig4}(a)].

To better track the FM-AFM transition of Fe$_x$Rh$_{100-x}$ films, the $M(T)$ and $\rho(T)$ were also collected 
upon heating and cooling the temperature in a field of $\mu_0$$H$ = 5\,T. 
For $33 \le x \le 53$, the $M(T)$ exhibits a significant drop below 400\,K upon 
cooling (see solid lines in the up panels in Fig.~\ref{fig:fig5}), indicating 
that these Fe$_x$Rh$_{100-x}$ films undergo a 
magnetic phase transition from high-$T$ FM state to 
low-$T$ AFM state. 
Such a FM-AFM transition is clearly reflected also in $\rho(T)$ data. As shown by solid lines in the bottom panels of 
Fig.~\ref{fig:fig5}, 
in contrast to the $M(T)$ data, the $\rho(T)$ undergoes a significant jump near the FM-AFM transition. In the FM state, 
both the Fe- and Rh moments are aligned along the $c$-axis~\cite{Shirane1963}, resulting in a low-resistivity state. 
While in the AFM state, 
the enhanced magnetic scattering leads to a high-resistivity state. 
When increasing the magnetic field, the Fe moments are forced to ferromagnetically align again,
accompanied by a resistivity drop at the metamagnetic transition~\cite{deVries_2013}. 
Upon heating, all the Fe$_x$Rh$_{100-x}$ films undergo an AFM-FM transition, reflected by a jump in the magnetization 
or a drop in the electrical resistivity (see dashed lines in 
Fig.~\ref{fig:fig5}).   
While for $x$ = 57 [see Fig.~\ref{fig:fig5}(f)], similar to the results in 
Fig.~\ref{fig:fig4}, no trace of magnetic 
transition can be identified in a field of 5\,T, consistent with its FM nature in the studied temperature range. 

\begin{figure}[!htp]
	\centering
	\includegraphics[width = 0.49\textwidth]{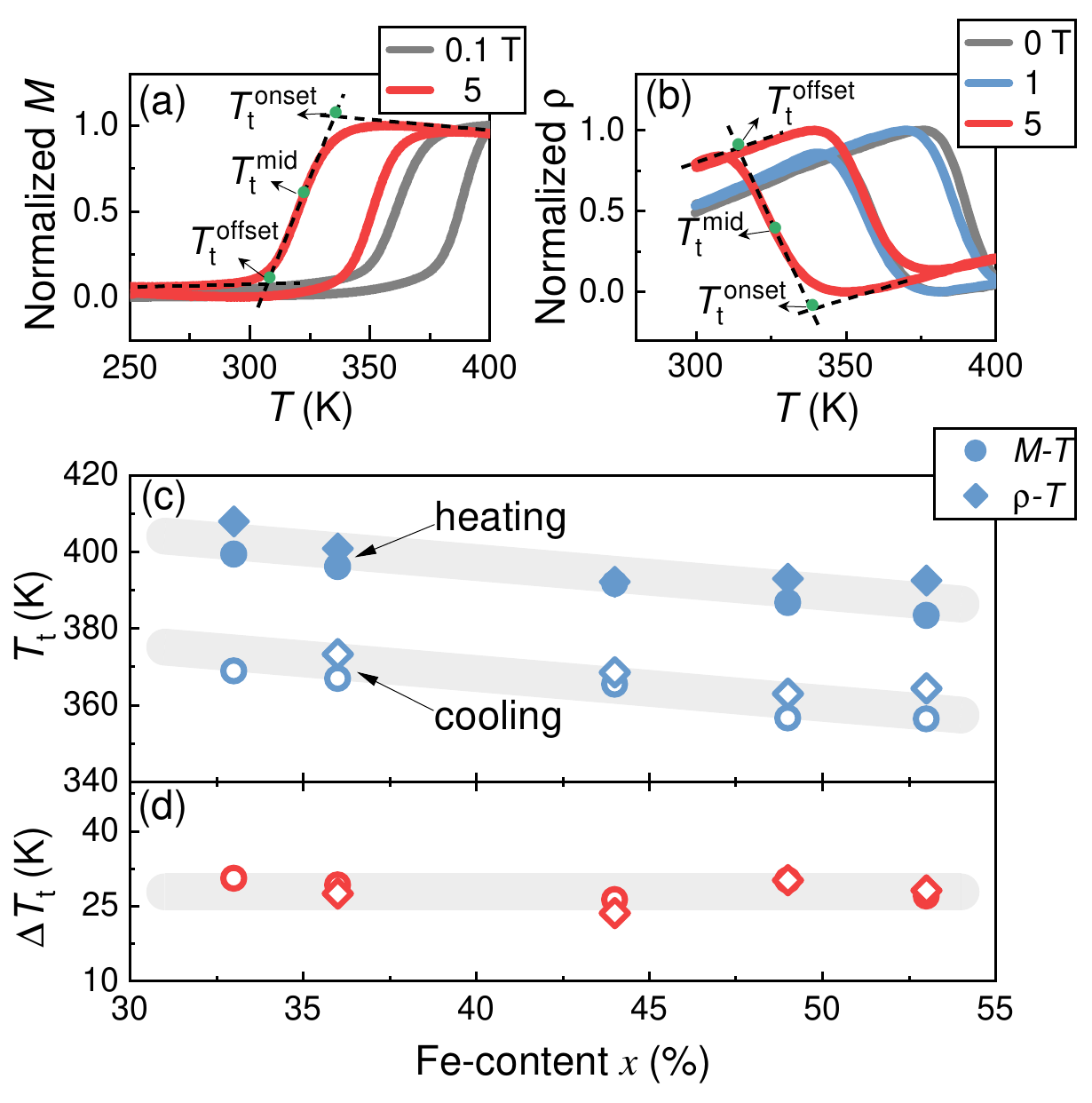}
	\vspace{-5ex}%
	\caption{\label{fig:fig6}\tcr{ Temperature-dependent magnetization 
	(a) and electrical resistivity (b) collected 
	under various magnetic fields up to 5\,T for Fe$_{49}$Rh$_{51}$ film. Both the magnetization and electrical resistivity data are 
	normalized to the values between 0 and 1.
		(c) The magnetic transition temperatures $T_\mathrm{t}$ (i.e., $T_\mathrm{t}^\mathrm{mid}$) determined from magnetization- and electrical-resistivity measurements vs. Fe-content. Open- and solid symbols represent the $T_\mathrm{t}$ determined from the measurements upon cooling- and heating processes, respectively. (d) The transition width $\Delta$$T_\mathrm{t}$ (= $T_\mathrm{t}^\mathrm{onset}$ - $T_\mathrm{t}^\mathrm{offset}$) of the FM-AFM (or AFM-FM) transition vs. the Fe-content.}}
\end{figure}

\begin{figure}[!htp]
	\centering
	\includegraphics[width = 0.48\textwidth]{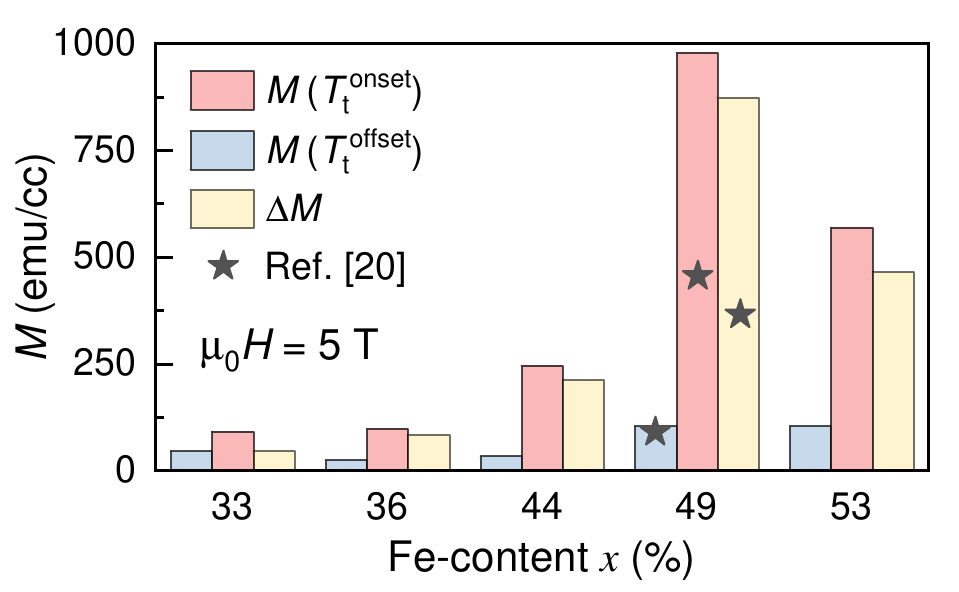}
	\vspace{-2ex}%
	\caption{\label{fig:fig7} The magnetization at $T_\mathrm{t}^\mathrm{onset}$ and $T_\mathrm{t}^\mathrm{offset}$, and 
	their difference 
		$\Delta$$M$ [= $M(T_\mathrm{t}^\mathrm{onset})$ - 
		$M(T_\mathrm{t}^\mathrm{offset})$] for various Fe$_x$Rh$_{100-x}$ films ($33 \le x \le 53$). 
		The data were obtained from the magnetization collected during the cooling process (see details in 
		Fig.~\ref{fig:fig4}). For the magnetization at $T_\mathrm{t}^\mathrm{onset}$ and $T_\mathrm{t}^\mathrm{offset}$, 
		the background signal from the MgO substrates was subtracted according to $M_\mathrm{FeRh}(T_\mathrm{t}^\mathrm{onset})$ = $M_\mathrm{raw}(T_\mathrm{t}^\mathrm{onset})$ - $M_\mathrm{MgO}(T_\mathrm{t}^\mathrm{onset})$.
		The data extracted from Ref.~\onlinecite{APL_THE} were also presented.}
\end{figure}

Figures~\ref{fig:fig6}(a) and \ref{fig:fig6}(b) plot the normalized 
temperature-dependent 
magnetization  and electrical resistivity collected under various magnetic 
fields up to 5\,T between 250 and 400\,K for 
Fe$_{49}$Rh$_{51}$ film. 
When increasing the magnetic field, the FM-AFM (or AFM-FM) transition is suppressed to lower temperatures. 
As indicated by the arrows, 
the $T_\mathrm{t}^\mathrm{onset}$, $T_\mathrm{t}^\mathrm{mid}$, and 
$T_\mathrm{t}^\mathrm{offset}$ are defined as the onset, 
middle, and offset of the magnetic transition temperatures, respectively. 
We found that $T_\mathrm{t}$ is suppressed at a rate of -7\,K/T by the external magnetic field for Fe$_{49}$Rh$_{51}$ 
film. The $T_\mathrm{t}(H)$ exhibits a linear field dependence up to 10\,T~\cite{deVries_2013}. 
To obtain zero-field magnetic transition temperatures, $T_\mathrm{t}$ values 
determined from 5\,T-data (see Fig.~\ref{fig:fig5}) were extrapolated to zero 
field using the above rate.    
The estimated zero-field  $T_\mathrm{t}$ values (here we choose 
$T_\mathrm{t}^\mathrm{mid}$) are summarized in Fig.~\ref{fig:fig6}(c) as a 
function of Fe-content.
The $T_\mathrm{t}$ determined from magnetization- and electrical-resistivity 
measurements are highly consistent. When increasing the Fe-content, 
$T_\mathrm{t}$ determined during the heating process slightly decreases from 
400\,K for $x$ = 33 to 385\,K for $x$ = 53. The $T_\mathrm{t}$ determined during 
the cooling process exhibits an almost identical trend, yielding a
$x$-independent transition width $\Delta$$T_t$ [see Fig.~\ref{fig:fig6}(d)]. 
Such $\Delta$$T_t(x)$ indicates that the first-order FM-AFM transition 
exists in the Fe$_{x}$Rh$_{100-x}$ ($33 \le x \le 53$) films with a much wider 
Fe-content than the bulk alloys.   
For the latter case, it is limited only at $48 \le x \le 54$ (see details in 
Fig.~\ref{fig:fig1}).

To quantitatively describe the FM-AFM transition in Fe$_{x}$Rh$_{100-x}$ ($33 \le x \le 53$) films, their magnetization at 
different magnetic states are summarized in Fig.~\ref{fig:fig7}. The
$M(T_\mathrm{t}^\mathrm{onset})$ represents the magnetization at 
$T_\mathrm{t}^\mathrm{onset}$ (i.e., FM state), while 
$M(T_\mathrm{t}^\mathrm{offset})$ is the magnetization at 
$T_\mathrm{t}^\mathrm{offset}$ (i.e., AFM state).
Both $M(T_\mathrm{t}^\mathrm{onset})$ and 
$M(T_\mathrm{t}^\mathrm{offset})$ reach a maximum value as the 
Fe-content increases up to 49\%.
According to the XRD results [see Fig.~\ref{fig:fig2}(a)], all the Fe$_{x}$Rh$_{100-x}$ ($33 \le x \le 53$) films show a pure 
$\alpha$-phase at room temperature, which is completely different from the bulk alloys (see Fig.~\ref{fig:fig1}). 
Therefore, in the Fe$_{x}$Rh$_{100-x}$ 
films, the larger magnetization 
value indicates a larger FM phase concentration at $T \ge T_\mathrm{t}^\mathrm{onset}$, and vice versa. As can be clearly 
seen in Fig.~\ref{fig:fig7}, the magnetization of $x$ = 49 and 53 films is 
significantly larger than that of $x < 49$, 
implying that most of the Fe moments stay PM in the latter cases. 
In the AFM state (i.e., $T \le T_\mathrm{t}^\mathrm{offset}$), the magnetization is mainly attributed to the pinned Fe moments at the Ta/Fe$_{x}$Rh$_{100-x}$ or Fe$_{x}$Rh$_{100-x}$/MgO interfaces~\cite{Fan_2010,Loving_2013}. 
As a consequence, the smaller magnetization value indicates a larger AFM phase concentration at $T \le T_\mathrm{t}^\mathrm{offset}$, and vice versa.
We also summarized the magnetization drop $\Delta$$M$, a measure of the FM-AFM 
transition, versus the Fe-content in Fig.~\ref{fig:fig6}. 
Similar to the $M(T_\mathrm{t}^\mathrm{onset})$, the $\Delta$$M$ also reaches a maximum value at $x$ = 49, which is 
significantly larger than the rest of Fe$_{x}$Rh$_{100-x}$ films.

For the Fe-Rh-based spintronic applications, 
the Fe$_{x}$Rh$_{100-x}$ films with a large $\Delta$$M$ value are preferred~\cite{Marti_2014,Thiele_2004,QIAO_2020}, since it could also give rise to a pronounced jump in the electrical resistivity.
The estimated resistivity jumps $\Delta$$\rho$ [= 
$\rho(T_\mathrm{t}^\mathrm{offset})$ - $\rho(T_\mathrm{t}^\mathrm{onset})$] 
[see details in Fig.~\ref{fig:fig6}(b)] of Fe$_{x}$Rh$_{100-x}$ ($33 \le x \le 
53$) films are summarized in the phase diagram (see Fig.~\ref{fig:fig1}). 
Indeed, for $x$ = 49 and 53, the $\Delta$$\rho$ values are significantly larger 
than that of $x < 49$. For instance, the $\Delta$$\rho$ = 
55~\textmu$\mathrm{\Omega}$cm for $x$ = 49, while it is less than 
10~\textmu$\mathrm{\Omega}$cm  for $x$ = 33.
Our results demonstrate that the Fe$_{x}$Rh$_{100-x}$ films with 
Fe-content up to 53\% exhibit magnetic and transport 
properties that are comparable to the  ideal 49\%-case.
While for $x \le 44$, the FM-AFM transition is less pronounced, leading to small 
$\Delta$$M$ and $\Delta$$\rho$ values.   

\begin{figure*}[!htp]
	\centering
	\includegraphics[width = 1\textwidth]{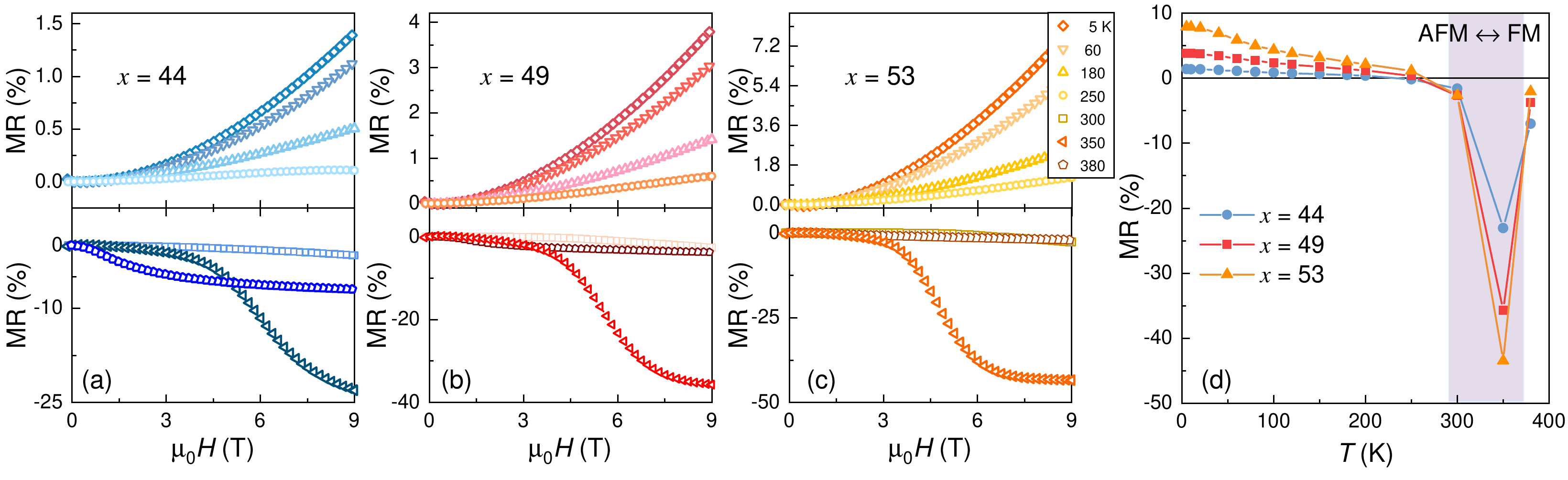}
	\vspace{-2ex}%
	\caption{\label{fig:fig8} Magnetoresistivity up to 9\,T for (a) $x = 44$, (b) $x=49$, and (c) $ x = 53$ collected at 
	various temperatures covering both the FM- and AFM states. 
		 (d) Temperature-dependent 9-T MR values for the above 
		 three films.   	
		The magnetic field was applied along the out-of-plane direction.
		The MR was calculated following MR = [$\rho(H)$ - 
		$\rho(0)$]/$\rho(0)$, where $\rho(0)$ is the zero-field electrical resistivity. 
		The shaded region highlights the coexistence of AFM- and FM phases, where a giant MR appears.}
\end{figure*}

\subsection{Magnetoresistivity and Hall resistivity}
%
\begin{figure}[!htp]
	\centering
	\includegraphics[width = 0.48\textwidth]{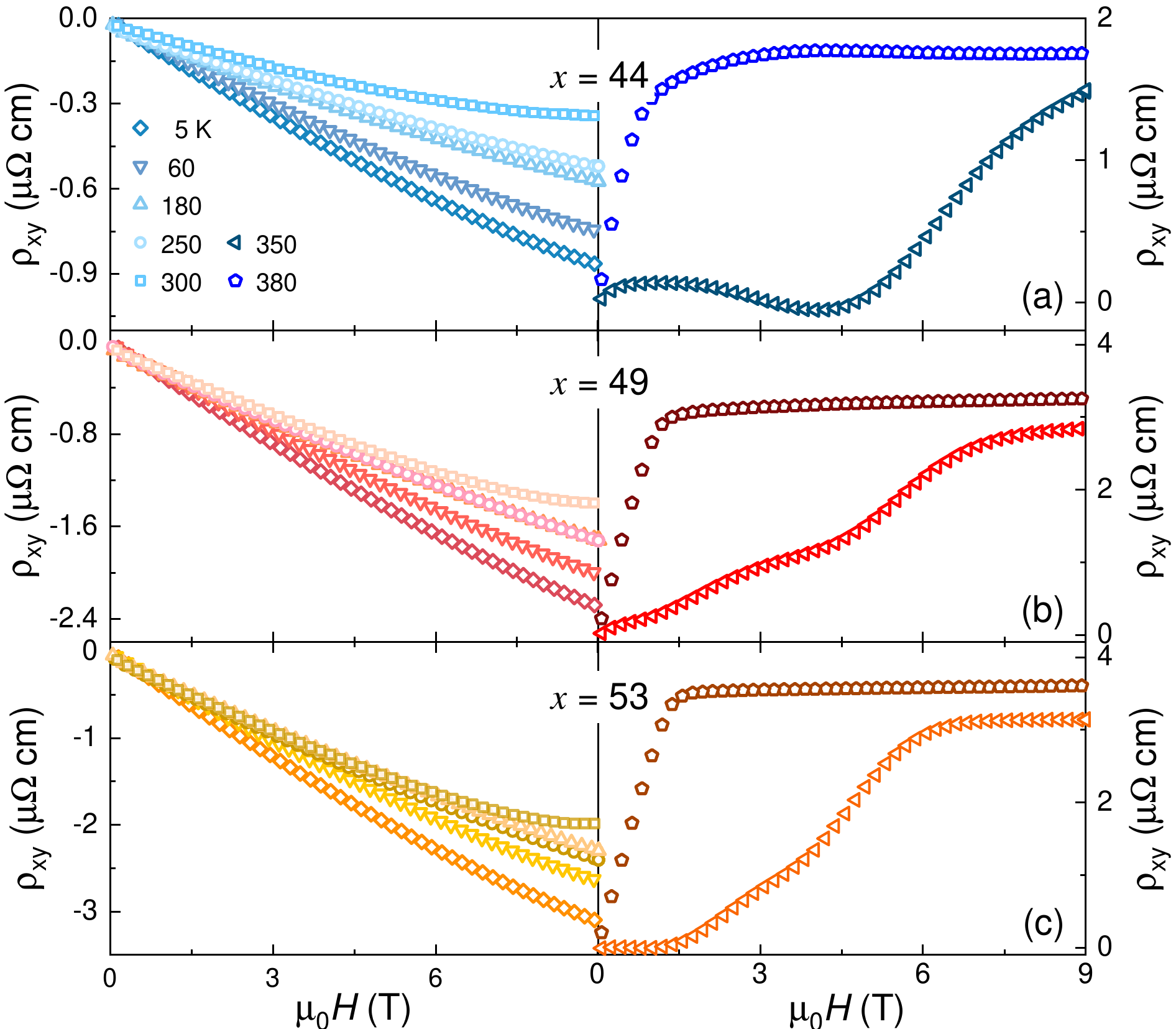}
	\vspace{-2ex}%
	\caption{\label{fig:fig9} Field-dependent Hall resistivity $\rho_\mathrm{xy}(H)$ collected at various temperature 
	below 400\,K up to 9\,T for Fe$_{x}$Rh$_{100-x}$ films with (a) $x = 44$, (b) $x$ = 49, and (c) $x$ = 53.
	}
\end{figure}

The field-dependent longitudinal- and transverse resistivity were measured in a wide temperature range for Fe$_{x}$Rh$_{100-x}$ films.
Since the films with $x$ = 44, 49, and 53 exhibit a more pronounced magnetic phase transition, 
here, the field-dependent measurements were focused on these films. 
Figure~\ref{fig:fig8}(a)-(c) plot the magnetoresistivity (MR) collected at 
various temperatures with the  magnetic field up to 
9\,T. The MR values of Fe$_{x}$Rh$_{100-x}$ films at $\mu_0H$ = 9\,T are 
summarized in Fig.~\ref{fig:fig8}(d).
All three films exhibit similar temperature-dependent MR in the 
studied temperature range. 
In the AFM state ($T \le 250$ K), the MR is positive, 
which is mainly attributed to the enhanced   
magnetic scattering by applying external magnetic field. 
Once the magnetic field destroys the AFM phase and fully polarizes the Fe moments, 
the MR exhibits a significant drop near the 
metamagnetic transition~\cite{deVries_2013,suzuki_2011}. 
As the temperature increases close to room temperature, where 
the FM phase 
shows up, the MR becomes negative, 
 a typical feature for the ferromagnets. 
While in the mixed AFM- and FM states, a giant MR was observed, whose value reaching almost 50\% at $T$ = 350\,K for $x$ = 53. 
Such a giant MR is related to the field-induced metamagnetic transition in Fe$_{x}$Rh$_{100-x}$ films, as observed in their AFM state~\cite{suzuki_2011}.  

The Fe$_{50}$Rh$_{50}$ film has been found to exhibit
a topological Hall effect in a wide temperature range~\cite{APL_THE}, which is often attributed to the topological spin textures in magnetic 
materials~\cite{raju_evolution_2019,Zadorozhnyi_2023}. Since the Fe-Rh alloys have a simple G-type AFM structure, the appearance of THE is rather puzzling.
To further investigate the possible THE in Fe$_{x}$Rh$_{100-x}$ films, we also performed systematic Hall-resistivity measurements. 
As shown in Fig.~\ref{fig:fig8}, the $\rho_\mathrm{xy}(H)$ were collected at various temperatures covering both the AFM- and FM states for Fe$_{x}$Rh$_{100-x}$ ($x$ = 44, 49, and 53) films. 
In the FM state, the $\rho_\mathrm{xy}(H)$ is dominated by the anomalous Hall effect 
(see 380-\,K curves in Fig.~\ref{fig:fig9}).    
While in the AFM state (i.e., $T <$ 300\,K), in contrast to the previous work~\cite{APL_THE},
all the $\rho_\mathrm{xy}(H)$ curves exhibit almost a linear field dependence at $\mu_0H \le$ 6\,T, definitely excluding the 
possible THE in our Fe$_{x}$Rh$_{100-x}$ films. While for $\mu_0H >$ 6\,T, the $\rho_\mathrm{xy}$ 
becomes nonlinear, which is clearly reflected by the 300-\,K curves in 
Fig.~\ref{fig:fig9}. 
Such nonlinear $\rho_\mathrm{xy}(H)$ is attributed to the field-induced metamagnetic transition in Fe$_{x}$Rh$_{100-x}$ films. 
The metamagnetic transition field is about 8.3\,T at 300\,K, which increases when decreasing temperature, reaching 9.8\,T at $T$ = 291\,K~\cite{deVries_2013}. 
Therefore, the $\rho_\mathrm{xy}$ is always dominated by the ordinary Hall effect (OHE) at $\mu_0H \le$ 9\,T for $T <$ 250\,K. 
However, once the magnetic field is larger than the metamagnetic transition field, 
the $\rho_\mathrm{xy}(H)$ resembles the typical features of AHE in 
ferromagnets. 
Interestingly, in the mixed AFM- and FM states, our Fe$_{x}$Rh$_{100-x}$ films exhibit a clear hump-like anomaly in the 
$\rho_\mathrm{xy}(H)$. As shown in Fig.~\ref{fig:fig9},
a clear hump can be observed at $\mu_0H \sim$ 3\,T at 350\,K. Such a hump-like anomaly resembles the topological Hall resistivity reported in the previous work~\cite{APL_THE}. 
However, such an anomaly is absent in the AFM state, implying that its origin is very unlikely the noncollinear spin textures. 
On the contrary, this anomaly can be reproduced, assuming anomalous Hall resistivity with different origins existing in the Fe$_{x}$Rh$_{100-x}$ films 
(see details in the discussion section).

\subsection{Discussion}

First, we discuss the magnetic phase diagram of Fe$_{x}$Rh$_{100-x}$ films.
After optimizing the growth conditions, we could produce phase-pure Fe$_{x}$Rh$_{100-x}$ films with a wide $x$ range (i.e., Fe-content). 
For the bulk case, the Fe$_{x}$Rh$_{100-x}$ alloys adopt the mixed $\alpha$- and $\gamma$-phases for $33 \le x \le 48$ [see details in Fig.~\ref{fig:fig1}(a)]. 
While the $\gamma$-phase remains PM, the $\alpha$-phase becomes FM below certain temperatures.
For $x >$ 48, the Fe$_{x}$Rh$_{100-x}$ alloys show a pure $\alpha$-phase with the Curie temperatures between 600 and 1000\,K. For some particular Fe concentrations, i.e., $48 \le x \le 54$, the Fe$_{x}$Rh$_{100-x}$ alloys  
undergo multiple magnetic transitions, from PM state to the FM state, and then finally to the AFM state. 
Different from the bulk alloys, Fe$_{x}$Rh$_{100-x}$ films show significantly different structural and magnetic properties. 
For the Fe-deficient case, Fe$_{30}$Rh$_{70}$ film adopts a 
$\gamma$-phase, and there is no magnetic transition below 400\,K, implying its PM nature.  
For $33 \le x \le 57$, all Fe$_{x}$Rh$_{100-x}$ films show a pure $\alpha$-phase. 
For the bulk alloys, no additional magnetic transition 
has been found below the Curie temperature for $33 \le x \le 48$. 
While in the case of films, there is a distinct FM-AFM 
transition for $33 \le x \le 53$, whose critical temperature 
$T_\mathrm{t}$ determined during the heating process slightly decreases from  
400\,K for $x$ = 33 to 385\,K for $x$ = 53 [marked as $\alpha$-(AFM+PM) in 
Fig.~\ref{fig:fig1}(a)].  In the case of Fe-rich films (i.e., $x$ $\ge$ 57 ), 
though they show pure $\alpha$-phase, the FM-AFM transition is absent, and all 
the films host a FM ground state. It is noted that the FM-AFM transition exists 
in the Fe$_{x}$Rh$_{100-x}$ films with a wide Fe-content, however, there is a 
large portion of remaining PM phase for $x \le 44$, which is reflected by a 
reduced resistivity jump $\Delta\rho$ and a magnetization drop $\Delta$$M$ (see 
details in Fig.~\ref{fig:fig1} and Fig.~\ref{fig:fig7}). 
In addition, in all the Fe$_{x}$Rh$_{100-x}$ films, the Fe moments pinned at the interfaces also contribute to the 
magnetization in the AFM state and could lead to a hump-like anomaly in the Hall resistivity. The absence of AHE in the AFM state proves that our 
Fe$_{x}$Rh$_{100-x}$ films have negligible remaining FM contribution (see 
Fig.~\ref{fig:fig9}).

%
\begin{figure}[!ht]
	\centering
	\includegraphics[width = 0.48\textwidth]{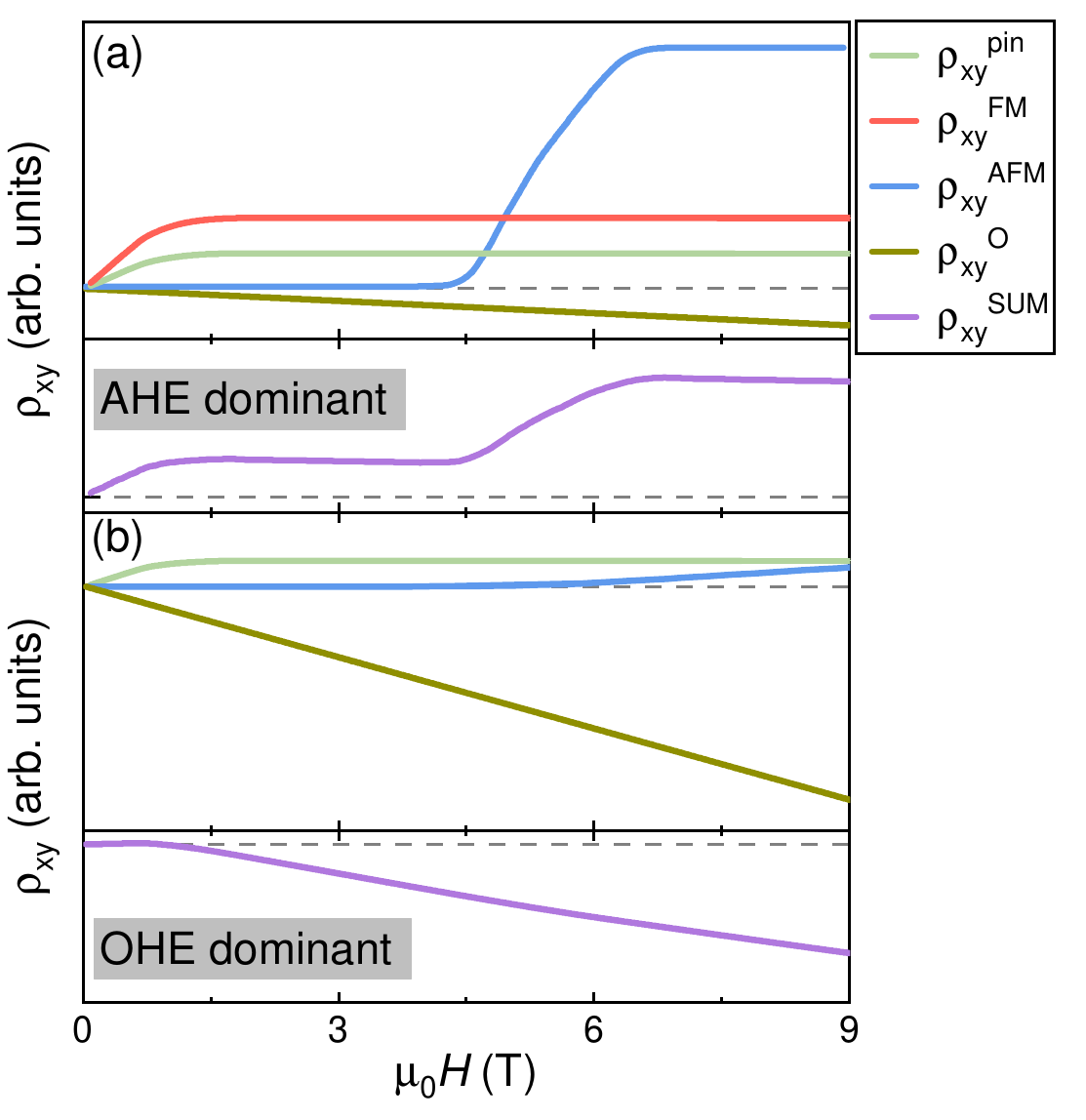}
	\vspace{-2ex}%
	\caption{\label{fig:fig10} \tcr{Schematic plots of different contributions to the Hall resistivity for the 
	Fe$_{x}$Rh$_{100-x}$ films 
		in the mixed AFM- and FM states (a) or in the AFM state 
		(b). In the mixed FM- and AFM states, 
		$\rho_\mathrm{xy}^\mathrm{SUM}$ =  
		$\rho_\mathrm{xy}^\mathrm{O}$ + 
		$\rho_\mathrm{xy}^\mathrm{pin}$ +  
		$\rho_\mathrm{xy}^\mathrm{FM}$
		+$\rho_\mathrm{xy}^\mathrm{AFM}$, while in the AFM state, 
		$\rho_\mathrm{xy}^\mathrm{SUM}$ =  
		$\rho_\mathrm{xy}^\mathrm{O}$ + 
		$\rho_\mathrm{xy}^\mathrm{pin}$ + 
		$\rho_\mathrm{xy}^\mathrm{AFM}$.
	    Here, $\rho_\mathrm{xy}^\mathrm{pin}$, $\rho_\mathrm{xy}^\mathrm{AFM}$, and $\rho_\mathrm{xy}^\mathrm{FM}$ all denote the anomalous-Hall resistivity.}}
\end{figure}

Now we discuss the possible THE in Fe$_{x}$Rh$_{100-x}$ films.  Based on 
the experimental observations in Fig.~\ref{fig:fig9}, we show the schematic 
plots in Fig.~\ref{fig:fig10} to discuss the Hall 
resistivity in Fe$_{x}$Rh$_{100-x}$ films. 
The linear $\rho_\mathrm{xy}(H)$ is caused by the OHE (marked as $\rho_\mathrm{xy}^\mathrm{O}$), whose 
negative slope suggests that the electron carriers are dominant in the 
Fe$_{x}$Rh$_{100-x}$ films.
The $\rho_\mathrm{xy}^\mathrm{pin}$, $\rho_\mathrm{xy}^\mathrm{AFM}$, and $\rho_\mathrm{xy}^\mathrm{FM}$ all denote the 
anomalous Hall resistivity, which are attributed to the ferromagnetically pinned 
Fe moments at the interfaces, AFM magnetization, and FM magnetization, respectively. 
In general, \tcr{both $\rho_\mathrm{xy}^\mathrm{pin}$ and 
$\rho_\mathrm{xy}^\mathrm{FM}$ are proportional to the 
magnetization (i.e., $\rho_\mathrm{xy}^\mathrm{pin}$ $\propto$ 
$M^\mathrm{pin}$, $\rho_\mathrm{xy}^\mathrm{FM}$  $\propto$ $M^\mathrm{FM}$}), 
typical for 
ferromagnets~\cite{Karplus_1954}. 
\tcr{Here, $M^\mathrm{pin}$ and $M^\mathrm{FM}$ are the magnetization attributed to the pinned Fe moments at the interfaces and the FM regions of the film, both of which saturate when 
	increasing the magnetic field up to 1\,T.}
	%
While the $\rho_\mathrm{xy}^\mathrm{AFM}$ is 
proportional to $\rho^2$$M$, $\rho$$M$, or their combinations, depending on the 
intrinsic or extrinsic mechanism~\cite{Nagaosa_2010}. Here we use $\rho$$M$ to 
produce
$\rho_\mathrm{xy}^\mathrm{AFM}$, while the $\rho^2$$M$ leads to similar 
	behaviors.  
\tcr{We assume that the AFM magnetization $M^\mathrm{AFM}$ is linear in the low-field region but undergoes a metamagnetic transition at higher magnetic field, whose critical field increases as lowering the temperature. For example, the metamagnetic transition field is close to 5\,T near room temperature but increases to 7.5\,T at 290\,K~\cite{deVries_2013}.} 
As shown in Fig.~\ref{fig:fig10}(a), in the 
mixed AFM- and FM states, the AHE due to FM and AFM magnetization (i.e., $\rho_\mathrm{xy}^\mathrm{FM}$ and $\rho_\mathrm{xy}^\mathrm{AFM}$) is 
dominant, and as a consequence, the total Hall resistivity
$\rho_\mathrm{xy}^\mathrm{SUM}$ shows a step-like feature, typical for the magnets that undergo a metamagentic transition.
The $\rho_\mathrm{xy}^\mathrm{SUM}$ qualitatively agrees very well with
$\rho_\mathrm{xy}(H)$ collected at 350\,K. 
In the AFM state, as shown in Fig.~\ref{fig:fig10}(b), since the 
OHE is dominant, the observed 
$\rho_\mathrm{xy}(H)$ is almost 
linear in field. While the $\rho_\mathrm{xy}^\mathrm{pin}$ could cause a hump-like anomaly in the $\rho_\mathrm{xy}^\mathrm{SUM}$, resembling 
the observed topological Hall resistivity in Ref.~[\onlinecite{APL_THE}]. 
However, such an anomaly is clearly absent in our Fe$_{x}$Rh$_{100-x}$ films 
with different Fe-contents (see Fig.~\ref{fig:fig9}). 
Since the $\rho_\mathrm{xy}^\mathrm{pin}$ is attributed to the FM iron moments pinned at the interfaces, such a hump-like 
anomaly in the $\rho_\mathrm{xy}(H)$ should strongly depend on the thin-film quality. 
We summarized the magnetization values from Ref.~[\onlinecite{APL_THE}] in 
Fig.~\ref{fig:fig7} to compare with our films. 
Though the remaining magnetization $M(T_\mathrm{t}^\mathrm{offset})$ is comparable to our 
Fe$_{49}$Rh$_{51}$ film, the FM-state magnetization $M(T_\mathrm{t}^\mathrm{onset})$ is two times smaller than our film. As 
a consequence, the $\rho_\mathrm{xy}(H)$ is significantly affected by the $\rho_\mathrm{xy}^\mathrm{pin}$ in the previous work. 
In addition, since the $\rho_\mathrm{xy}^\mathrm{pin}$ has opposite sign against the $\rho_\mathrm{xy}^\mathrm{O}$, 
the large contribution of $\rho_\mathrm{xy}^\mathrm{pin}$ might cause the sign change in the $\rho_\mathrm{xy}(H)$ when 
cooling the film down to lower temperatures. Indeed, such a sign change was observed in the previous work, the slope of $\rho_\mathrm{xy}(H)$ 
becomes positive below 80\,K~\cite{APL_THE}. While in our Fe$_{x}$Rh$_{100-x}$ ($x$ = 44, 49, and 53) films, the 
$\rho_\mathrm{xy}(H)$ is 
always negative in the AFM state, which again proves that the remaining FM magnetization at the interfaces has little effect 
in our Fe$_{x}$Rh$_{100-x}$ films. 
To conclude, the observed THE in Fe$_{x}$Rh$_{100-x}$ films 
is most likely an extrinsic effect. The other techniques, such as resonant x-ray scattering or Lorentz transmission 
electron microscopy, are highly desirable to search for possible topological magnetic phases in Fe$_{x}$Rh$_{100-x}$ family.

\section{CONCLUSION} 

To summarize, we grew a series of epitaxial Fe$_x$Rh$_{100-x}$ ($30 \le x \le 57$) films on MgO substrates. By systematic x-ray diffraction-, magnetization-, and electrical resistivity measurements, 
we established the structural and magnetic phase diagram of Fe$_x$Rh$_{100-x}$ films. 
For $x$ $\le$ 30, Fe$_x$Rh$_{100-x}$ films are PM and adopt a $\gamma$-phase. For $x 
\ge 33$, all films show a pure $\alpha$-phase.
In the bulk Fe$_x$Rh$_{100-x}$ alloys, the FM-AFM 
transition is limited only at $48 \leq x \leq 54$. While the FM-AFM 
transition persists in the Fe$_x$Rh$_{100-x}$ films with 
$33 \leq x \leq 53$, and the transition temperature slightly decreases from 
400\,K for $x$ = 33 to 385\,K for $x$ = 53. As further increases the 
Fe-content (i.e., $x > 53$), the FM-AFM transition no longer exists, and Fe$_x$Rh$_{100-x}$ films are FM in the studied 
temperature range. 
The resistivity jump and magnetization drop at the FM-AFM transition are much 
more pronounced in the Fe$_x$Rh$_{100-x}$ films with $\sim$50\% Fe-content 
than in the Fe-deficient films, the latter have a large amount of PM 
phase. The magnetoresistivity is positive and weak in the AFM state, while it 
becomes negative when the FM phase shows up, and a giant MR appears in the 
mixed AFM- and FM states. The Hall resistivity measurements reveal trivial 
behaviors in the Fe$_x$Rh$_{100-x}$ films, which is dominated by the OHE 
in the AFM state and by the AHE in the mixed- or 
FM state, respectively. Our results demonstrate that the previously observed 
topological Hall resistivity is absent in our Fe$_{x}$Rh$_{100-x}$ ($x$ = 44, 
49, and 53) films. We proposed that the AHE caused by the FM iron 
moments at the interfaces could explain the hump-like anomaly in the Hall 
resistivity. To conclude, the observed THE in 
Fe$_{x}$Rh$_{100-x}$ films can be explained by extrinsic mechanisms rather than the presence of noncollinear spin textures. 
%
%
\begin{acknowledgments}
	The authors thank G. T. Lin and J. Ma for their assistance during the transport measurements.  
	This work was supported by  
    the Natural Science Foundation of Shanghai 
	(Grants No.\ 21ZR1420500 and 21JC\-140\-2300), Natural Science
	Foundation of Chongqing (Grant No.\ 2022NSCQ-MSX1468), the National Natural Science Foundation of China (Grants No. 
	12174103 and 12374105).	Y.X.\ acknowledges support from the Shanghai Pujiang Program (Grant No.\ 21PJ1403100).
\end{acknowledgments}

\bibliography{FeRh.bib}

\begin{thebibliography}{41}%
\makeatletter
\providecommand \@ifxundefined [1]{%
 \@ifx{#1\undefined}
}%
\providecommand \@ifnum [1]{%
 \ifnum #1\expandafter \@firstoftwo
 \else \expandafter \@secondoftwo
 \fi
}%
\providecommand \@ifx [1]{%
 \ifx #1\expandafter \@firstoftwo
 \else \expandafter \@secondoftwo
 \fi
}%
\providecommand \natexlab [1]{#1}%
\providecommand \enquote  [1]{``#1''}%
\providecommand \bibnamefont  [1]{#1}%
\providecommand \bibfnamefont [1]{#1}%
\providecommand \citenamefont [1]{#1}%
\providecommand \href@noop [0]{\@secondoftwo}%
\providecommand \href [0]{\begingroup \@sanitize@url \@href}%
\providecommand \@href[1]{\@@startlink{#1}\@@href}%
\providecommand \@@href[1]{\endgroup#1\@@endlink}%
\providecommand \@sanitize@url [0]{\catcode `\\12\catcode `\$12\catcode
  `\&12\catcode `\#12\catcode `\^12\catcode `\_12\catcode `\%12\relax}%
\providecommand \@@startlink[1]{}%
\providecommand \@@endlink[0]{}%
\providecommand \url  [0]{\begingroup\@sanitize@url \@url }%
\providecommand \@url [1]{\endgroup\@href {#1}{\urlprefix }}%
\providecommand \urlprefix  [0]{URL }%
\providecommand \Eprint [0]{\href }%
\providecommand \doibase [0]{https://doi.org/}%
\providecommand \selectlanguage [0]{\@gobble}%
\providecommand \bibinfo  [0]{\@secondoftwo}%
\providecommand \bibfield  [0]{\@secondoftwo}%
\providecommand \translation [1]{[#1]}%
\providecommand \BibitemOpen [0]{}%
\providecommand \bibitemStop [0]{}%
\providecommand \bibitemNoStop [0]{.\EOS\space}%
\providecommand \EOS [0]{\spacefactor3000\relax}%
\providecommand \BibitemShut  [1]{\csname bibitem#1\endcsname}%
\let\auto@bib@innerbib\@empty
\bibitem [{\citenamefont {Uhl\'{l}\v{r}}\ \emph {et~al.}(2016)\citenamefont
  {Uhl\'{l}\v{r}}, \citenamefont {Arregi},\ and\ \citenamefont
  {Fullerton}}]{Uhl_2016}%
  \BibitemOpen
  \bibfield  {author} {\bibinfo {author} {\bibfnamefont {V.}~\bibnamefont
  {Uhl\'{l}\v{r}}}, \bibinfo {author} {\bibfnamefont {J.~A.}\ \bibnamefont
  {Arregi}},\ and\ \bibinfo {author} {\bibfnamefont {E.~E.}\ \bibnamefont
  {Fullerton}},\ }\bibfield  {title} {\bibinfo {title} {Colossal magnetic phase
  transition asymmetry in mesoscale {FeRh} stripes},\ }\href
  {https://doi.org/10.1038/ncomms13113} {\bibfield  {journal} {\bibinfo
  {journal} {Nat. Commun.}\ }\textbf {\bibinfo {volume} {7}},\ \bibinfo {pages}
  {13113} (\bibinfo {year} {2016})}\BibitemShut {NoStop}%
\bibitem [{\citenamefont {Li}\ \emph {et~al.}(2022)\citenamefont {Li},
  \citenamefont {Medapalli}, \citenamefont {Mentink}, \citenamefont
  {Mikhaylovskiy}, \citenamefont {Blank}, \citenamefont {Patel}, \citenamefont
  {Zvezdin}, \citenamefont {Rasing}, \citenamefont {Fullerton},\ and\
  \citenamefont {Kimel}}]{Li_2022}%
  \BibitemOpen
  \bibfield  {author} {\bibinfo {author} {\bibfnamefont {G.}~\bibnamefont
  {Li}}, \bibinfo {author} {\bibfnamefont {R.}~\bibnamefont {Medapalli}},
  \bibinfo {author} {\bibfnamefont {J.~H.}\ \bibnamefont {Mentink}}, \bibinfo
  {author} {\bibfnamefont {R.~V.}\ \bibnamefont {Mikhaylovskiy}}, \bibinfo
  {author} {\bibfnamefont {T.~G.~H.}\ \bibnamefont {Blank}}, \bibinfo {author}
  {\bibfnamefont {S.~K.~K.}\ \bibnamefont {Patel}}, \bibinfo {author}
  {\bibfnamefont {A.~K.}\ \bibnamefont {Zvezdin}}, \bibinfo {author}
  {\bibfnamefont {T.}~\bibnamefont {Rasing}}, \bibinfo {author} {\bibfnamefont
  {E.~E.}\ \bibnamefont {Fullerton}},\ and\ \bibinfo {author} {\bibfnamefont
  {A.~V.}\ \bibnamefont {Kimel}},\ }\bibfield  {title} {\bibinfo {title}
  {Ultrafast kinetics of the antiferromagnetic-ferromagnetic phase transition
  in {FeRh}},\ }\href {https://doi.org/10.1038/s41467-022-30591-2} {\bibfield
  {journal} {\bibinfo  {journal} {Nat. Commun.}\ }\textbf {\bibinfo {volume}
  {13}},\ \bibinfo {pages} {2998} (\bibinfo {year} {2022})}\BibitemShut
  {NoStop}%
\bibitem [{\citenamefont {Kinane}\ \emph {et~al.}(2014)\citenamefont {Kinane},
  \citenamefont {Loving}, \citenamefont {Vries}, \citenamefont {Fan},
  \citenamefont {Charlton}, \citenamefont {Claydon}, \citenamefont {Arena},
  \citenamefont {Maccherozzi}, \citenamefont {Dhesi}, \citenamefont {Heiman},
  \citenamefont {Marrows}, \citenamefont {Lewis},\ and\ \citenamefont
  {Langridge}}]{Kinane_2014}%
  \BibitemOpen
  \bibfield  {author} {\bibinfo {author} {\bibfnamefont {C.~J.}\ \bibnamefont
  {Kinane}}, \bibinfo {author} {\bibfnamefont {M.}~\bibnamefont {Loving}},
  \bibinfo {author} {\bibfnamefont {M.~A.~d.}\ \bibnamefont {Vries}}, \bibinfo
  {author} {\bibfnamefont {R.}~\bibnamefont {Fan}}, \bibinfo {author}
  {\bibfnamefont {T.~R.}\ \bibnamefont {Charlton}}, \bibinfo {author}
  {\bibfnamefont {J.~S.}\ \bibnamefont {Claydon}}, \bibinfo {author}
  {\bibfnamefont {D.~A.}\ \bibnamefont {Arena}}, \bibinfo {author}
  {\bibfnamefont {F.}~\bibnamefont {Maccherozzi}}, \bibinfo {author}
  {\bibfnamefont {S.~S.}\ \bibnamefont {Dhesi}}, \bibinfo {author}
  {\bibfnamefont {D.}~\bibnamefont {Heiman}}, \bibinfo {author} {\bibfnamefont
  {C.~H.}\ \bibnamefont {Marrows}}, \bibinfo {author} {\bibfnamefont {L.~H.}\
  \bibnamefont {Lewis}},\ and\ \bibinfo {author} {\bibfnamefont
  {S.}~\bibnamefont {Langridge}},\ }\bibfield  {title} {\bibinfo {title}
  {Observation of a temperature dependent asymmetry in the domain structure of
  a {Pd}-doped {FeRh} epilayer},\ }\href
  {https://doi.org/10.1088/1367-2630/16/11/113073} {\bibfield  {journal}
  {\bibinfo  {journal} {New J. Phys.}\ }\textbf {\bibinfo {volume} {16}},\
  \bibinfo {pages} {113073} (\bibinfo {year} {2014})}\BibitemShut {NoStop}%
\bibitem [{\citenamefont {Arregi}\ \emph {et~al.}(2020)\citenamefont {Arregi},
  \citenamefont {Caha},\ and\ \citenamefont {Uhl\'{\i}\ifmmode~\check{r}\else
  \v{r}\fi{}}}]{Arregi_2020}%
  \BibitemOpen
  \bibfield  {author} {\bibinfo {author} {\bibfnamefont {J.~A.}\ \bibnamefont
  {Arregi}}, \bibinfo {author} {\bibfnamefont {O.~c.~v.}\ \bibnamefont
  {Caha}},\ and\ \bibinfo {author} {\bibfnamefont {V.~c.~v.}\ \bibnamefont
  {Uhl\'{\i}\ifmmode~\check{r}\else \v{r}\fi{}}},\ }\bibfield  {title}
  {\bibinfo {title} {Evolution of strain across the magnetostructural phase
  transition in epitaxial ferh films on different substrates},\ }\href
  {https://doi.org/10.1103/PhysRevB.101.174413} {\bibfield  {journal} {\bibinfo
   {journal} {Phys. Rev. B}\ }\textbf {\bibinfo {volume} {101}},\ \bibinfo
  {pages} {174413} (\bibinfo {year} {2020})}\BibitemShut {NoStop}%
\bibitem [{\citenamefont {Xie}\ \emph {et~al.}(2017)\citenamefont {Xie},
  \citenamefont {Zhan}, \citenamefont {Shang}, \citenamefont {Yang},
  \citenamefont {Wang}, \citenamefont {Tang},\ and\ \citenamefont
  {Li}}]{Xie_2017}%
  \BibitemOpen
  \bibfield  {author} {\bibinfo {author} {\bibfnamefont {Y.}~\bibnamefont
  {Xie}}, \bibinfo {author} {\bibfnamefont {Q.}~\bibnamefont {Zhan}}, \bibinfo
  {author} {\bibfnamefont {T.}~\bibnamefont {Shang}}, \bibinfo {author}
  {\bibfnamefont {H.}~\bibnamefont {Yang}}, \bibinfo {author} {\bibfnamefont
  {B.}~\bibnamefont {Wang}}, \bibinfo {author} {\bibfnamefont {J.}~\bibnamefont
  {Tang}},\ and\ \bibinfo {author} {\bibfnamefont {R.-W.}\ \bibnamefont {Li}},\
  }\bibfield  {title} {\bibinfo {title} {Effect of epitaxial strain and lattice
  mismatch on magnetic and transport behaviors in metamagnetic {FeRh} thin
  films},\ }\href {https://doi.org/10.1063/1.4976301} {\bibfield  {journal}
  {\bibinfo  {journal} {AIP Adv.}\ }\textbf {\bibinfo {volume} {7}},\ \bibinfo
  {pages} {056314} (\bibinfo {year} {2017})}\BibitemShut {NoStop}%
\bibitem [{\citenamefont {Qiao}\ \emph {et~al.}(2019)\citenamefont {Qiao},
  \citenamefont {Hu}, \citenamefont {Liu}, \citenamefont {Li}, \citenamefont
  {Kuang}, \citenamefont {Zhang}, \citenamefont {Liang}, \citenamefont {Wang},
  \citenamefont {Sun},\ and\ \citenamefont {Shen}}]{QIAO_2019}%
  \BibitemOpen
  \bibfield  {author} {\bibinfo {author} {\bibfnamefont {K.}~\bibnamefont
  {Qiao}}, \bibinfo {author} {\bibfnamefont {F.}~\bibnamefont {Hu}}, \bibinfo
  {author} {\bibfnamefont {Y.}~\bibnamefont {Liu}}, \bibinfo {author}
  {\bibfnamefont {J.}~\bibnamefont {Li}}, \bibinfo {author} {\bibfnamefont
  {H.}~\bibnamefont {Kuang}}, \bibinfo {author} {\bibfnamefont
  {H.}~\bibnamefont {Zhang}}, \bibinfo {author} {\bibfnamefont
  {W.}~\bibnamefont {Liang}}, \bibinfo {author} {\bibfnamefont
  {J.}~\bibnamefont {Wang}}, \bibinfo {author} {\bibfnamefont {J.}~\bibnamefont
  {Sun}},\ and\ \bibinfo {author} {\bibfnamefont {B.}~\bibnamefont {Shen}},\
  }\bibfield  {title} {\bibinfo {title} {Novel reduction of hysteresis loss
  controlled by strain memory effect in {FeRh}/{PMN-PT} heterostructures},\
  }\href {https://doi.org/https://doi.org/10.1016/j.nanoen.2019.02.044}
  {\bibfield  {journal} {\bibinfo  {journal} {Nano Energy}\ }\textbf {\bibinfo
  {volume} {59}},\ \bibinfo {pages} {285} (\bibinfo {year} {2019})}\BibitemShut
  {NoStop}%
\bibitem [{\citenamefont {Cherifi}\ \emph {et~al.}(2014)\citenamefont
  {Cherifi}, \citenamefont {Ivanovskaya}, \citenamefont {Phillips},
  \citenamefont {Zobelli}, \citenamefont {Infante}, \citenamefont {Jacquet},
  \citenamefont {Garcia}, \citenamefont {Fusil}, \citenamefont {Briddon},
  \citenamefont {Guiblin}, \citenamefont {Mougin}, \citenamefont {\"{U}nal},
  \citenamefont {Kronast}, \citenamefont {Valencia}, \citenamefont {Dkhil},
  \citenamefont {Barth\'{e}l\'{e}my},\ and\ \citenamefont
  {Bibes}}]{Cherifi_2014}%
  \BibitemOpen
  \bibfield  {author} {\bibinfo {author} {\bibfnamefont {R.~O.}\ \bibnamefont
  {Cherifi}}, \bibinfo {author} {\bibfnamefont {V.}~\bibnamefont
  {Ivanovskaya}}, \bibinfo {author} {\bibfnamefont {L.~C.}\ \bibnamefont
  {Phillips}}, \bibinfo {author} {\bibfnamefont {A.}~\bibnamefont {Zobelli}},
  \bibinfo {author} {\bibfnamefont {I.~C.}\ \bibnamefont {Infante}}, \bibinfo
  {author} {\bibfnamefont {E.}~\bibnamefont {Jacquet}}, \bibinfo {author}
  {\bibfnamefont {V.}~\bibnamefont {Garcia}}, \bibinfo {author} {\bibfnamefont
  {S.}~\bibnamefont {Fusil}}, \bibinfo {author} {\bibfnamefont {P.~R.}\
  \bibnamefont {Briddon}}, \bibinfo {author} {\bibfnamefont {N.}~\bibnamefont
  {Guiblin}}, \bibinfo {author} {\bibfnamefont {A.}~\bibnamefont {Mougin}},
  \bibinfo {author} {\bibfnamefont {A.~A.}\ \bibnamefont {\"{U}nal}}, \bibinfo
  {author} {\bibfnamefont {F.}~\bibnamefont {Kronast}}, \bibinfo {author}
  {\bibfnamefont {S.}~\bibnamefont {Valencia}}, \bibinfo {author}
  {\bibfnamefont {B.}~\bibnamefont {Dkhil}}, \bibinfo {author} {\bibfnamefont
  {A.}~\bibnamefont {Barth\'{e}l\'{e}my}},\ and\ \bibinfo {author}
  {\bibfnamefont {M.}~\bibnamefont {Bibes}},\ }\bibfield  {title} {\bibinfo
  {title} {Electric-field control of magnetic order above room temperature},\
  }\href {https://doi.org/10.1038/nmat3870} {\bibfield  {journal} {\bibinfo
  {journal} {Nat. Mater.}\ }\textbf {\bibinfo {volume} {13}},\ \bibinfo {pages}
  {345} (\bibinfo {year} {2014})}\BibitemShut {NoStop}%
\bibitem [{\citenamefont {Maat}\ \emph {et~al.}(2005)\citenamefont {Maat},
  \citenamefont {Thiele},\ and\ \citenamefont {Fullerton}}]{Maat_2005}%
  \BibitemOpen
  \bibfield  {author} {\bibinfo {author} {\bibfnamefont {S.}~\bibnamefont
  {Maat}}, \bibinfo {author} {\bibfnamefont {J.-U.}\ \bibnamefont {Thiele}},\
  and\ \bibinfo {author} {\bibfnamefont {E.~E.}\ \bibnamefont {Fullerton}},\
  }\bibfield  {title} {\bibinfo {title} {Temperature and field hysteresis of
  the antiferromagnetic-to-ferromagnetic phase transition in epitaxial {FeRh}
  films},\ }\href {https://doi.org/10.1103/PhysRevB.72.214432} {\bibfield
  {journal} {\bibinfo  {journal} {Phys. Rev. B}\ }\textbf {\bibinfo {volume}
  {72}},\ \bibinfo {pages} {214432} (\bibinfo {year} {2005})}\BibitemShut
  {NoStop}%
\bibitem [{\citenamefont {Lee}\ \emph {et~al.}(2015)\citenamefont {Lee},
  \citenamefont {Liu}, \citenamefont {Heron}, \citenamefont {Clarkson},
  \citenamefont {Hong}, \citenamefont {Ko}, \citenamefont {Biegalski},
  \citenamefont {Aschauer}, \citenamefont {Hsu}, \citenamefont {Nowakowski},
  \citenamefont {Wu}, \citenamefont {Christen}, \citenamefont {Salahuddin},
  \citenamefont {Bokor}, \citenamefont {Spaldin}, \citenamefont {Schlom},\ and\
  \citenamefont {Ramesh}}]{Lee_2015}%
  \BibitemOpen
  \bibfield  {author} {\bibinfo {author} {\bibfnamefont {Y.}~\bibnamefont
  {Lee}}, \bibinfo {author} {\bibfnamefont {Z.~Q.}\ \bibnamefont {Liu}},
  \bibinfo {author} {\bibfnamefont {J.~T.}\ \bibnamefont {Heron}}, \bibinfo
  {author} {\bibfnamefont {J.~D.}\ \bibnamefont {Clarkson}}, \bibinfo {author}
  {\bibfnamefont {J.}~\bibnamefont {Hong}}, \bibinfo {author} {\bibfnamefont
  {C.}~\bibnamefont {Ko}}, \bibinfo {author} {\bibfnamefont {M.~D.}\
  \bibnamefont {Biegalski}}, \bibinfo {author} {\bibfnamefont {U.}~\bibnamefont
  {Aschauer}}, \bibinfo {author} {\bibfnamefont {S.~L.}\ \bibnamefont {Hsu}},
  \bibinfo {author} {\bibfnamefont {M.~E.}\ \bibnamefont {Nowakowski}},
  \bibinfo {author} {\bibfnamefont {J.}~\bibnamefont {Wu}}, \bibinfo {author}
  {\bibfnamefont {H.~M.}\ \bibnamefont {Christen}}, \bibinfo {author}
  {\bibfnamefont {S.}~\bibnamefont {Salahuddin}}, \bibinfo {author}
  {\bibfnamefont {J.~B.}\ \bibnamefont {Bokor}}, \bibinfo {author}
  {\bibfnamefont {N.~A.}\ \bibnamefont {Spaldin}}, \bibinfo {author}
  {\bibfnamefont {D.~G.}\ \bibnamefont {Schlom}},\ and\ \bibinfo {author}
  {\bibfnamefont {R.}~\bibnamefont {Ramesh}},\ }\bibfield  {title} {\bibinfo
  {title} {Large resistivity modulation in mixed-phase metallic systems},\
  }\href {https://doi.org/10.1038/ncomms6959} {\bibfield  {journal} {\bibinfo
  {journal} {Nat. Commun.}\ }\textbf {\bibinfo {volume} {6}},\ \bibinfo {pages}
  {5959} (\bibinfo {year} {2015})}\BibitemShut {NoStop}%
\bibitem [{\citenamefont {Liu}\ \emph {et~al.}(2016{\natexlab{a}})\citenamefont
  {Liu}, \citenamefont {Li}, \citenamefont {Gai}, \citenamefont {Clarkson},
  \citenamefont {Hsu}, \citenamefont {Wong}, \citenamefont {Fan}, \citenamefont
  {Lin}, \citenamefont {Rouleau}, \citenamefont {Ward}, \citenamefont {Lee},
  \citenamefont {Sefat}, \citenamefont {Christen},\ and\ \citenamefont
  {Ramesh}}]{Liu_2016}%
  \BibitemOpen
  \bibfield  {author} {\bibinfo {author} {\bibfnamefont {Z.~Q.}\ \bibnamefont
  {Liu}}, \bibinfo {author} {\bibfnamefont {L.}~\bibnamefont {Li}}, \bibinfo
  {author} {\bibfnamefont {Z.}~\bibnamefont {Gai}}, \bibinfo {author}
  {\bibfnamefont {J.~D.}\ \bibnamefont {Clarkson}}, \bibinfo {author}
  {\bibfnamefont {S.~L.}\ \bibnamefont {Hsu}}, \bibinfo {author} {\bibfnamefont
  {A.~T.}\ \bibnamefont {Wong}}, \bibinfo {author} {\bibfnamefont {L.~S.}\
  \bibnamefont {Fan}}, \bibinfo {author} {\bibfnamefont {M.-W.}\ \bibnamefont
  {Lin}}, \bibinfo {author} {\bibfnamefont {C.~M.}\ \bibnamefont {Rouleau}},
  \bibinfo {author} {\bibfnamefont {T.~Z.}\ \bibnamefont {Ward}}, \bibinfo
  {author} {\bibfnamefont {H.~N.}\ \bibnamefont {Lee}}, \bibinfo {author}
  {\bibfnamefont {A.~S.}\ \bibnamefont {Sefat}}, \bibinfo {author}
  {\bibfnamefont {H.~M.}\ \bibnamefont {Christen}},\ and\ \bibinfo {author}
  {\bibfnamefont {R.}~\bibnamefont {Ramesh}},\ }\bibfield  {title} {\bibinfo
  {title} {Full electroresistance modulation in a mixed-phase metallic alloy},\
  }\href {https://doi.org/10.1103/PhysRevLett.116.097203} {\bibfield  {journal}
  {\bibinfo  {journal} {Phys. Rev. Lett.}\ }\textbf {\bibinfo {volume} {116}},\
  \bibinfo {pages} {097203} (\bibinfo {year} {2016}{\natexlab{a}})}\BibitemShut
  {NoStop}%
\bibitem [{\citenamefont {Cao}\ \emph {et~al.}(2022)\citenamefont {Cao},
  \citenamefont {Chen}, \citenamefont {Cui}, \citenamefont {Yu}, \citenamefont
  {Jiang}, \citenamefont {Yang}, \citenamefont {Qiu}, \citenamefont {Shang},
  \citenamefont {Xu},\ and\ \citenamefont {Zhan}}]{Cao_2022}%
  \BibitemOpen
  \bibfield  {author} {\bibinfo {author} {\bibfnamefont {C.}~\bibnamefont
  {Cao}}, \bibinfo {author} {\bibfnamefont {S.}~\bibnamefont {Chen}}, \bibinfo
  {author} {\bibfnamefont {B.}~\bibnamefont {Cui}}, \bibinfo {author}
  {\bibfnamefont {G.}~\bibnamefont {Yu}}, \bibinfo {author} {\bibfnamefont
  {C.}~\bibnamefont {Jiang}}, \bibinfo {author} {\bibfnamefont
  {Z.}~\bibnamefont {Yang}}, \bibinfo {author} {\bibfnamefont {X.}~\bibnamefont
  {Qiu}}, \bibinfo {author} {\bibfnamefont {T.}~\bibnamefont {Shang}}, \bibinfo
  {author} {\bibfnamefont {Y.}~\bibnamefont {Xu}},\ and\ \bibinfo {author}
  {\bibfnamefont {Q.}~\bibnamefont {Zhan}},\ }\bibfield  {title} {\bibinfo
  {title} {Efficient {Tuning} of the {Spin}–{Orbit} {Torque} via the
  {Magnetic} {Phase} {Transition} of {FeRh}},\ }\href
  {https://doi.org/10.1021/acsnano.2c04488} {\bibfield  {journal} {\bibinfo
  {journal} {ACS Nano}\ }\textbf {\bibinfo {volume} {16}},\ \bibinfo {pages}
  {12727} (\bibinfo {year} {2022})}\BibitemShut {NoStop}%
\bibitem [{\citenamefont {Qiao}\ \emph {et~al.}(2020)\citenamefont {Qiao},
  \citenamefont {Wang}, \citenamefont {Hu}, \citenamefont {Li}, \citenamefont
  {Zhang}, \citenamefont {Liu}, \citenamefont {Yu}, \citenamefont {Gao},
  \citenamefont {Su}, \citenamefont {Shen}, \citenamefont {Zhou}, \citenamefont
  {Bai}, \citenamefont {Wang}, \citenamefont {Franco}, \citenamefont {Sun},\
  and\ \citenamefont {Shen}}]{QIAO_2020}%
  \BibitemOpen
  \bibfield  {author} {\bibinfo {author} {\bibfnamefont {K.}~\bibnamefont
  {Qiao}}, \bibinfo {author} {\bibfnamefont {J.}~\bibnamefont {Wang}}, \bibinfo
  {author} {\bibfnamefont {F.}~\bibnamefont {Hu}}, \bibinfo {author}
  {\bibfnamefont {J.}~\bibnamefont {Li}}, \bibinfo {author} {\bibfnamefont
  {C.}~\bibnamefont {Zhang}}, \bibinfo {author} {\bibfnamefont
  {Y.}~\bibnamefont {Liu}}, \bibinfo {author} {\bibfnamefont {Z.}~\bibnamefont
  {Yu}}, \bibinfo {author} {\bibfnamefont {Y.}~\bibnamefont {Gao}}, \bibinfo
  {author} {\bibfnamefont {J.}~\bibnamefont {Su}}, \bibinfo {author}
  {\bibfnamefont {F.}~\bibnamefont {Shen}}, \bibinfo {author} {\bibfnamefont
  {H.}~\bibnamefont {Zhou}}, \bibinfo {author} {\bibfnamefont {X.}~\bibnamefont
  {Bai}}, \bibinfo {author} {\bibfnamefont {J.}~\bibnamefont {Wang}}, \bibinfo
  {author} {\bibfnamefont {V.}~\bibnamefont {Franco}}, \bibinfo {author}
  {\bibfnamefont {J.}~\bibnamefont {Sun}},\ and\ \bibinfo {author}
  {\bibfnamefont {B.}~\bibnamefont {Shen}},\ }\bibfield  {title} {\bibinfo
  {title} {Regulation of phase transition and magnetocaloric effect by
  ferroelectric domains in {FeRh}/{PMN}-{PT} heterojunctions},\ }\href
  {https://doi.org/10.1016/j.actamat.2020.03.028} {\bibfield  {journal}
  {\bibinfo  {journal} {Acta Mater.}\ }\textbf {\bibinfo {volume} {191}},\
  \bibinfo {pages} {51} (\bibinfo {year} {2020})}\BibitemShut {NoStop}%
\bibitem [{\citenamefont {Marti}\ \emph {et~al.}(2014)\citenamefont {Marti},
  \citenamefont {Fina}, \citenamefont {Frontera}, \citenamefont {Liu},
  \citenamefont {Wadley}, \citenamefont {He}, \citenamefont {Paull},
  \citenamefont {Clarkson}, \citenamefont {Kudrnovsk\'{y}}, \citenamefont
  {Turek}, \citenamefont {Kune\v{s}}, \citenamefont {Yi}, \citenamefont {Chu},
  \citenamefont {Nelson}, \citenamefont {You}, \citenamefont {Arenholz},
  \citenamefont {Salahuddin}, \citenamefont {Fontcuberta}, \citenamefont
  {Jungwirth},\ and\ \citenamefont {Ramesh}}]{Marti_2014}%
  \BibitemOpen
  \bibfield  {author} {\bibinfo {author} {\bibfnamefont {X.}~\bibnamefont
  {Marti}}, \bibinfo {author} {\bibfnamefont {I.}~\bibnamefont {Fina}},
  \bibinfo {author} {\bibfnamefont {C.}~\bibnamefont {Frontera}}, \bibinfo
  {author} {\bibfnamefont {J.}~\bibnamefont {Liu}}, \bibinfo {author}
  {\bibfnamefont {P.}~\bibnamefont {Wadley}}, \bibinfo {author} {\bibfnamefont
  {Q.}~\bibnamefont {He}}, \bibinfo {author} {\bibfnamefont {R.~J.}\
  \bibnamefont {Paull}}, \bibinfo {author} {\bibfnamefont {J.~D.}\ \bibnamefont
  {Clarkson}}, \bibinfo {author} {\bibfnamefont {J.}~\bibnamefont
  {Kudrnovsk\'{y}}}, \bibinfo {author} {\bibfnamefont {I.}~\bibnamefont
  {Turek}}, \bibinfo {author} {\bibfnamefont {J.}~\bibnamefont {Kune\v{s}}},
  \bibinfo {author} {\bibfnamefont {D.}~\bibnamefont {Yi}}, \bibinfo {author}
  {\bibfnamefont {J.-H.}\ \bibnamefont {Chu}}, \bibinfo {author} {\bibfnamefont
  {C.~T.}\ \bibnamefont {Nelson}}, \bibinfo {author} {\bibfnamefont
  {L.}~\bibnamefont {You}}, \bibinfo {author} {\bibfnamefont {E.}~\bibnamefont
  {Arenholz}}, \bibinfo {author} {\bibfnamefont {S.}~\bibnamefont
  {Salahuddin}}, \bibinfo {author} {\bibfnamefont {J.}~\bibnamefont
  {Fontcuberta}}, \bibinfo {author} {\bibfnamefont {T.}~\bibnamefont
  {Jungwirth}},\ and\ \bibinfo {author} {\bibfnamefont {R.}~\bibnamefont
  {Ramesh}},\ }\bibfield  {title} {\bibinfo {title} {Room-temperature
  antiferromagnetic memory resistor},\ }\href
  {https://doi.org/10.1038/nmat3861} {\bibfield  {journal} {\bibinfo  {journal}
  {Nat. Mater.}\ }\textbf {\bibinfo {volume} {13}},\ \bibinfo {pages} {367}
  (\bibinfo {year} {2014})}\BibitemShut {NoStop}%
\bibitem [{\citenamefont {Thiele}\ \emph {et~al.}(2004)\citenamefont {Thiele},
  \citenamefont {Maat}, \citenamefont {Robertson},\ and\ \citenamefont
  {Fullerton}}]{Thiele_2004}%
  \BibitemOpen
  \bibfield  {author} {\bibinfo {author} {\bibfnamefont {J.-U.}\ \bibnamefont
  {Thiele}}, \bibinfo {author} {\bibfnamefont {S.}~\bibnamefont {Maat}},
  \bibinfo {author} {\bibfnamefont {J.~L.}\ \bibnamefont {Robertson}},\ and\
  \bibinfo {author} {\bibfnamefont {E.~E.}\ \bibnamefont {Fullerton}},\
  }\bibfield  {title} {\bibinfo {title} {Magnetic and {Structural} {Properties}
  of {FePt}–{FeRh} {Exchange} {Spring} {Films} for {Thermally} {Assisted}
  {Magnetic} {Recording} {Media}},\ }\href
  {https://doi.org/10.1109/TMAG.2004.829325} {\bibfield  {journal} {\bibinfo
  {journal} {IEEE Trans. Magn.}\ }\textbf {\bibinfo {volume} {40}},\ \bibinfo
  {pages} {2537} (\bibinfo {year} {2004})}\BibitemShut {NoStop}%
\bibitem [{\citenamefont {Bertaut}\ \emph {et~al.}(1962)\citenamefont
  {Bertaut}, \citenamefont {Delapalme}, \citenamefont {Forrat}, \citenamefont
  {Roult}, \citenamefont {De~Bergevin},\ and\ \citenamefont
  {Pauthenet}}]{Bertaut1962}%
  \BibitemOpen
  \bibfield  {author} {\bibinfo {author} {\bibfnamefont {E.~F.}\ \bibnamefont
  {Bertaut}}, \bibinfo {author} {\bibfnamefont {A.}~\bibnamefont {Delapalme}},
  \bibinfo {author} {\bibfnamefont {F.}~\bibnamefont {Forrat}}, \bibinfo
  {author} {\bibfnamefont {G.}~\bibnamefont {Roult}}, \bibinfo {author}
  {\bibfnamefont {F.}~\bibnamefont {De~Bergevin}},\ and\ \bibinfo {author}
  {\bibfnamefont {R.}~\bibnamefont {Pauthenet}},\ }\bibfield  {title} {\bibinfo
  {title} {Magnetic {Structure} {Work} at the {Nuclear} {Center} of
  {Grenoble}},\ }\href {https://doi.org/10.1063/1.1728627} {\bibfield
  {journal} {\bibinfo  {journal} {J. Appl. Phys.}\ }\textbf {\bibinfo {volume}
  {33}},\ \bibinfo {pages} {1123} (\bibinfo {year} {1962})}\BibitemShut
  {NoStop}%
\bibitem [{\citenamefont {Shirane}\ \emph
  {et~al.}(1963{\natexlab{a}})\citenamefont {Shirane}, \citenamefont {Chen},
  \citenamefont {Flinn},\ and\ \citenamefont {Nathans}}]{Shirane1963}%
  \BibitemOpen
  \bibfield  {author} {\bibinfo {author} {\bibfnamefont {G.}~\bibnamefont
  {Shirane}}, \bibinfo {author} {\bibfnamefont {C.~W.}\ \bibnamefont {Chen}},
  \bibinfo {author} {\bibfnamefont {P.~A.}\ \bibnamefont {Flinn}},\ and\
  \bibinfo {author} {\bibfnamefont {R.}~\bibnamefont {Nathans}},\ }\bibfield
  {title} {\bibinfo {title} {Hyperfine {Fields} and {Magnetic} {Moments} in the
  {Fe}–{Rh} {System}},\ }\href {https://doi.org/10.1063/1.1729362} {\bibfield
   {journal} {\bibinfo  {journal} {J. Appl. Phys.}\ }\textbf {\bibinfo {volume}
  {34}},\ \bibinfo {pages} {1044} (\bibinfo {year}
  {1963}{\natexlab{a}})}\BibitemShut {NoStop}%
\bibitem [{\citenamefont {Zhu}\ \emph {et~al.}(2022{\natexlab{a}})\citenamefont
  {Zhu}, \citenamefont {Xu}, \citenamefont {Cao}, \citenamefont {Shang},
  \citenamefont {Xie},\ and\ \citenamefont {Zhan}}]{Zhu_2022JPCM}%
  \BibitemOpen
  \bibfield  {author} {\bibinfo {author} {\bibfnamefont {X.}~\bibnamefont
  {Zhu}}, \bibinfo {author} {\bibfnamefont {Y.}~\bibnamefont {Xu}}, \bibinfo
  {author} {\bibfnamefont {C.}~\bibnamefont {Cao}}, \bibinfo {author}
  {\bibfnamefont {T.}~\bibnamefont {Shang}}, \bibinfo {author} {\bibfnamefont
  {Y.}~\bibnamefont {Xie}},\ and\ \bibinfo {author} {\bibfnamefont
  {Q.}~\bibnamefont {Zhan}},\ }\bibfield  {title} {\bibinfo {title} {Recent
  developments on the magnetic and electrical transport properties of {FeRh}-
  and {Rh}-based heterostructures},\ }\href
  {https://doi.org/10.1088/1361-648X/ac4b28} {\bibfield  {journal} {\bibinfo
  {journal} {J. Phys.: Condes. Matter}\ }\textbf {\bibinfo {volume} {34}},\
  \bibinfo {pages} {144004} (\bibinfo {year} {2022}{\natexlab{a}})}\BibitemShut
  {NoStop}%
\bibitem [{\citenamefont {Zhu}\ \emph {et~al.}(2022{\natexlab{b}})\citenamefont
  {Zhu}, \citenamefont {Li}, \citenamefont {Xie}, \citenamefont {Qiu},
  \citenamefont {Cao}, \citenamefont {Hu}, \citenamefont {Xie}, \citenamefont
  {Shang}, \citenamefont {Xu}, \citenamefont {Sun}, \citenamefont {Cheng},
  \citenamefont {Jiang},\ and\ \citenamefont {Zhan}}]{ZHU_2022JAC}%
  \BibitemOpen
  \bibfield  {author} {\bibinfo {author} {\bibfnamefont {X.}~\bibnamefont
  {Zhu}}, \bibinfo {author} {\bibfnamefont {Y.}~\bibnamefont {Li}}, \bibinfo
  {author} {\bibfnamefont {Y.}~\bibnamefont {Xie}}, \bibinfo {author}
  {\bibfnamefont {Q.}~\bibnamefont {Qiu}}, \bibinfo {author} {\bibfnamefont
  {C.}~\bibnamefont {Cao}}, \bibinfo {author} {\bibfnamefont {X.}~\bibnamefont
  {Hu}}, \bibinfo {author} {\bibfnamefont {W.}~\bibnamefont {Xie}}, \bibinfo
  {author} {\bibfnamefont {T.}~\bibnamefont {Shang}}, \bibinfo {author}
  {\bibfnamefont {Y.}~\bibnamefont {Xu}}, \bibinfo {author} {\bibfnamefont
  {L.}~\bibnamefont {Sun}}, \bibinfo {author} {\bibfnamefont {W.}~\bibnamefont
  {Cheng}}, \bibinfo {author} {\bibfnamefont {D.}~\bibnamefont {Jiang}},\ and\
  \bibinfo {author} {\bibfnamefont {Q.}~\bibnamefont {Zhan}},\ }\bibfield
  {title} {\bibinfo {title} {Magnetocrystalline anisotropy of epitaxially grown
  {FeRh}/{MgO}(001) films},\ }\href
  {https://doi.org/10.1016/j.jallcom.2022.165566} {\bibfield  {journal}
  {\bibinfo  {journal} {J. Alloy. Compd.}\ }\textbf {\bibinfo {volume} {917}},\
  \bibinfo {pages} {165566} (\bibinfo {year} {2022}{\natexlab{b}})}\BibitemShut
  {NoStop}%
\bibitem [{\citenamefont {Liu}\ \emph {et~al.}(2016{\natexlab{b}})\citenamefont
  {Liu}, \citenamefont {Phillips}, \citenamefont {Mattana}, \citenamefont
  {Bibes}, \citenamefont {Barth\'{e}l\'{e}my},\ and\ \citenamefont
  {Dkhil}}]{Liu_large_2016}%
  \BibitemOpen
  \bibfield  {author} {\bibinfo {author} {\bibfnamefont {Y.}~\bibnamefont
  {Liu}}, \bibinfo {author} {\bibfnamefont {L.~C.}\ \bibnamefont {Phillips}},
  \bibinfo {author} {\bibfnamefont {R.}~\bibnamefont {Mattana}}, \bibinfo
  {author} {\bibfnamefont {M.}~\bibnamefont {Bibes}}, \bibinfo {author}
  {\bibfnamefont {A.}~\bibnamefont {Barth\'{e}l\'{e}my}},\ and\ \bibinfo
  {author} {\bibfnamefont {B.}~\bibnamefont {Dkhil}},\ }\bibfield  {title}
  {\bibinfo {title} {Large reversible caloric effect in {FeRh} thin films via a
  dual-stimulus multicaloric cycle},\ }\href
  {https://doi.org/10.1038/ncomms11614} {\bibfield  {journal} {\bibinfo
  {journal} {Nat. Commun.}\ }\textbf {\bibinfo {volume} {7}},\ \bibinfo {pages}
  {11614} (\bibinfo {year} {2016}{\natexlab{b}})}\BibitemShut {NoStop}%
\bibitem [{\citenamefont {Zhang}\ \emph {et~al.}(2019)\citenamefont {Zhang},
  \citenamefont {Xia}, \citenamefont {Cao}, \citenamefont {Wang}, \citenamefont
  {Liu},\ and\ \citenamefont {Du}}]{APL_THE}%
  \BibitemOpen
  \bibfield  {author} {\bibinfo {author} {\bibfnamefont {S.}~\bibnamefont
  {Zhang}}, \bibinfo {author} {\bibfnamefont {S.}~\bibnamefont {Xia}}, \bibinfo
  {author} {\bibfnamefont {Q.}~\bibnamefont {Cao}}, \bibinfo {author}
  {\bibfnamefont {D.}~\bibnamefont {Wang}}, \bibinfo {author} {\bibfnamefont
  {R.}~\bibnamefont {Liu}},\ and\ \bibinfo {author} {\bibfnamefont
  {Y.}~\bibnamefont {Du}},\ }\bibfield  {title} {\bibinfo {title} {Observation
  of topological {Hall} effect in antiferromagnetic {FeRh} film},\ }\href
  {https://doi.org/10.1063/1.5099183} {\bibfield  {journal} {\bibinfo
  {journal} {Appl. Phys. Lett.}\ }\textbf {\bibinfo {volume} {115}},\ \bibinfo
  {pages} {022404} (\bibinfo {year} {2019})}\BibitemShut {NoStop}%
\bibitem [{\citenamefont {Matsuno}\ \emph {et~al.}(2016)\citenamefont
  {Matsuno}, \citenamefont {Ogawa}, \citenamefont {Yasuda}, \citenamefont
  {Kagawa}, \citenamefont {Koshibae}, \citenamefont {Nagaosa}, \citenamefont
  {Tokura},\ and\ \citenamefont {Kawasaki}}]{Matsuno_interface-driven_2016}%
  \BibitemOpen
  \bibfield  {author} {\bibinfo {author} {\bibfnamefont {J.}~\bibnamefont
  {Matsuno}}, \bibinfo {author} {\bibfnamefont {N.}~\bibnamefont {Ogawa}},
  \bibinfo {author} {\bibfnamefont {K.}~\bibnamefont {Yasuda}}, \bibinfo
  {author} {\bibfnamefont {F.}~\bibnamefont {Kagawa}}, \bibinfo {author}
  {\bibfnamefont {W.}~\bibnamefont {Koshibae}}, \bibinfo {author}
  {\bibfnamefont {N.}~\bibnamefont {Nagaosa}}, \bibinfo {author} {\bibfnamefont
  {Y.}~\bibnamefont {Tokura}},\ and\ \bibinfo {author} {\bibfnamefont
  {M.}~\bibnamefont {Kawasaki}},\ }\bibfield  {title} {\bibinfo {title}
  {Interface-driven topological {Hall} effect in
  {SrRuO}$_{\textrm{3}}$-{SrIrO}$_{\textrm{3}}$ bilayer},\ }\href
  {https://doi.org/10.1126/sciadv.1600304} {\bibfield  {journal} {\bibinfo
  {journal} {Sci. Adv.}\ }\textbf {\bibinfo {volume} {2}},\ \bibinfo {pages}
  {e1600304} (\bibinfo {year} {2016})}\BibitemShut {NoStop}%
\bibitem [{\citenamefont {Neubauer}\ \emph {et~al.}(2009)\citenamefont
  {Neubauer}, \citenamefont {Pfleiderer}, \citenamefont {Binz}, \citenamefont
  {Rosch}, \citenamefont {Ritz}, \citenamefont {Niklowitz},\ and\ \citenamefont
  {B\"{o}ni}}]{Neubauer_2009}%
  \BibitemOpen
  \bibfield  {author} {\bibinfo {author} {\bibfnamefont {A.}~\bibnamefont
  {Neubauer}}, \bibinfo {author} {\bibfnamefont {C.}~\bibnamefont
  {Pfleiderer}}, \bibinfo {author} {\bibfnamefont {B.}~\bibnamefont {Binz}},
  \bibinfo {author} {\bibfnamefont {A.}~\bibnamefont {Rosch}}, \bibinfo
  {author} {\bibfnamefont {R.}~\bibnamefont {Ritz}}, \bibinfo {author}
  {\bibfnamefont {P.~G.}\ \bibnamefont {Niklowitz}},\ and\ \bibinfo {author}
  {\bibfnamefont {P.}~\bibnamefont {B\"{o}ni}},\ }\bibfield  {title} {\bibinfo
  {title} {Topological {Hall} {Effect} in the $a$ {Phase} of {MnSi}},\ }\href
  {https://doi.org/10.1103/PhysRevLett.102.186602} {\bibfield  {journal}
  {\bibinfo  {journal} {Phys. Rev. Lett.}\ }\textbf {\bibinfo {volume} {102}},\
  \bibinfo {pages} {186602} (\bibinfo {year} {2009})}\BibitemShut {NoStop}%
\bibitem [{\citenamefont {Kimbell}\ \emph {et~al.}(2022)\citenamefont
  {Kimbell}, \citenamefont {Kim}, \citenamefont {Wu}, \citenamefont {Cuoco},\
  and\ \citenamefont {Robinson}}]{FAKE_THE}%
  \BibitemOpen
  \bibfield  {author} {\bibinfo {author} {\bibfnamefont {G.}~\bibnamefont
  {Kimbell}}, \bibinfo {author} {\bibfnamefont {C.}~\bibnamefont {Kim}},
  \bibinfo {author} {\bibfnamefont {W.}~\bibnamefont {Wu}}, \bibinfo {author}
  {\bibfnamefont {M.}~\bibnamefont {Cuoco}},\ and\ \bibinfo {author}
  {\bibfnamefont {J.~W.~A.}\ \bibnamefont {Robinson}},\ }\bibfield  {title}
  {\bibinfo {title} {Challenges in identifying chiral spin textures via the
  topological {Hall} effect},\ }\href
  {https://doi.org/10.1038/s43246-022-00238-2} {\bibfield  {journal} {\bibinfo
  {journal} {Commun. Mater.}\ }\textbf {\bibinfo {volume} {3}},\ \bibinfo
  {pages} {19} (\bibinfo {year} {2022})}\BibitemShut {NoStop}%
\bibitem [{\citenamefont {Groenendijk}\ \emph {et~al.}(2020)\citenamefont
  {Groenendijk}, \citenamefont {Autieri}, \citenamefont {van Thiel},
  \citenamefont {Brzezicki}, \citenamefont {Hortensius}, \citenamefont
  {Afanasiev}, \citenamefont {Gauquelin}, \citenamefont {Barone}, \citenamefont
  {van~den Bos}, \citenamefont {van Aert}, \citenamefont {Verbeeck},
  \citenamefont {Filippetti}, \citenamefont {Picozzi}, \citenamefont {Cuoco},\
  and\ \citenamefont {Caviglia}}]{Fake_THE2}%
  \BibitemOpen
  \bibfield  {author} {\bibinfo {author} {\bibfnamefont {D.~J.}\ \bibnamefont
  {Groenendijk}}, \bibinfo {author} {\bibfnamefont {C.}~\bibnamefont
  {Autieri}}, \bibinfo {author} {\bibfnamefont {T.~C.}\ \bibnamefont {van
  Thiel}}, \bibinfo {author} {\bibfnamefont {W.}~\bibnamefont {Brzezicki}},
  \bibinfo {author} {\bibfnamefont {J.~R.}\ \bibnamefont {Hortensius}},
  \bibinfo {author} {\bibfnamefont {D.}~\bibnamefont {Afanasiev}}, \bibinfo
  {author} {\bibfnamefont {N.}~\bibnamefont {Gauquelin}}, \bibinfo {author}
  {\bibfnamefont {P.}~\bibnamefont {Barone}}, \bibinfo {author} {\bibfnamefont
  {K.~H.~W.}\ \bibnamefont {van~den Bos}}, \bibinfo {author} {\bibfnamefont
  {S.}~\bibnamefont {van Aert}}, \bibinfo {author} {\bibfnamefont
  {J.}~\bibnamefont {Verbeeck}}, \bibinfo {author} {\bibfnamefont
  {A.}~\bibnamefont {Filippetti}}, \bibinfo {author} {\bibfnamefont
  {S.}~\bibnamefont {Picozzi}}, \bibinfo {author} {\bibfnamefont
  {M.}~\bibnamefont {Cuoco}},\ and\ \bibinfo {author} {\bibfnamefont {A.~D.}\
  \bibnamefont {Caviglia}},\ }\bibfield  {title} {\bibinfo {title} {Berry phase
  engineering at oxide interfaces},\ }\href
  {https://doi.org/10.1103/PhysRevResearch.2.023404} {\bibfield  {journal}
  {\bibinfo  {journal} {Phys. Rev. Res.}\ }\textbf {\bibinfo {volume} {2}},\
  \bibinfo {pages} {023404} (\bibinfo {year} {2020})}\BibitemShut {NoStop}%
\bibitem [{\citenamefont {Wang}\ \emph {et~al.}(2020)\citenamefont {Wang},
  \citenamefont {Feng}, \citenamefont {Lee}, \citenamefont {Ko}, \citenamefont
  {Lu},\ and\ \citenamefont {Noh}}]{FAKE_THE3}%
  \BibitemOpen
  \bibfield  {author} {\bibinfo {author} {\bibfnamefont {L.}~\bibnamefont
  {Wang}}, \bibinfo {author} {\bibfnamefont {Q.}~\bibnamefont {Feng}}, \bibinfo
  {author} {\bibfnamefont {H.~G.}\ \bibnamefont {Lee}}, \bibinfo {author}
  {\bibfnamefont {E.~K.}\ \bibnamefont {Ko}}, \bibinfo {author} {\bibfnamefont
  {Q.}~\bibnamefont {Lu}},\ and\ \bibinfo {author} {\bibfnamefont {T.~W.}\
  \bibnamefont {Noh}},\ }\bibfield  {title} {\bibinfo {title} {Controllable
  {Thickness} {Inhomogeneity} and {Berry} {Curvature} {Engineering} of
  {Anomalous} {Hall} {Effect} in {SrRuO}$_{\textrm{3}}$ {Ultrathin} {Films}},\
  }\href {https://doi.org/10.1021/acs.nanolett.9b05206} {\bibfield  {journal}
  {\bibinfo  {journal} {Nano Lett.}\ }\textbf {\bibinfo {volume} {20}},\
  \bibinfo {pages} {2468} (\bibinfo {year} {2020})}\BibitemShut {NoStop}%
\bibitem [{\citenamefont {Zhang}\ and\ \citenamefont
  {Granville}(2022)}]{Zhang_2022}%
  \BibitemOpen
  \bibfield  {author} {\bibinfo {author} {\bibfnamefont {Y.}~\bibnamefont
  {Zhang}}\ and\ \bibinfo {author} {\bibfnamefont {S.}~\bibnamefont
  {Granville}},\ }\bibfield  {title} {\bibinfo {title} {Two-channel anomalous
  hall effect originating from the intermixing in {Mn}$_2${CoAl}/{Pd} thin
  films},\ }\href {https://doi.org/10.1103/PhysRevB.106.144414} {\bibfield
  {journal} {\bibinfo  {journal} {Phys. Rev. B}\ }\textbf {\bibinfo {volume}
  {106}},\ \bibinfo {pages} {144414} (\bibinfo {year} {2022})}\BibitemShut
  {NoStop}%
\bibitem [{\citenamefont {Goldbeck}(1982)}]{1982Iron}%
  \BibitemOpen
  \bibfield  {author} {\bibinfo {author} {\bibfnamefont {O.~K.}\ \bibnamefont
  {Goldbeck}},\ }\href@noop {} {\emph {\bibinfo {title} {{IRON}-{Binary}
  {Phase} {Diagrams}}}}\ (\bibinfo  {publisher} {Springer Berlin Heidelberg},\
  \bibinfo {address} {Berlin, Heidelberg},\ \bibinfo {year} {1982})\BibitemShut
  {NoStop}%
\bibitem [{\citenamefont {Inoue}\ \emph {et~al.}(2008)\citenamefont {Inoue},
  \citenamefont {Yu~Yu~Ko},\ and\ \citenamefont {Suzuki}}]{Inoue_2008}%
  \BibitemOpen
  \bibfield  {author} {\bibinfo {author} {\bibfnamefont {S.}~\bibnamefont
  {Inoue}}, \bibinfo {author} {\bibfnamefont {H.}~\bibnamefont {Yu~Yu~Ko}},\
  and\ \bibinfo {author} {\bibfnamefont {T.}~\bibnamefont {Suzuki}},\
  }\bibfield  {title} {\bibinfo {title} {Magnetic {Properties} of
  {Single}-{Crystalline} {FeRh} {Alloy} {Thin} {Films}},\ }\href
  {https://doi.org/10.1109/TMAG.2008.2001846} {\bibfield  {journal} {\bibinfo
  {journal} {IEEE Trans. Magn.}\ }\textbf {\bibinfo {volume} {44}},\ \bibinfo
  {pages} {2875} (\bibinfo {year} {2008})}\BibitemShut {NoStop}%
\bibitem [{\citenamefont {Mei}\ \emph {et~al.}(2018)\citenamefont {Mei},
  \citenamefont {Tang}, \citenamefont {Grab}, \citenamefont {Schubert},
  \citenamefont {Ralph},\ and\ \citenamefont {Schlom}}]{Mei_2018}%
  \BibitemOpen
  \bibfield  {author} {\bibinfo {author} {\bibfnamefont {A.~B.}\ \bibnamefont
  {Mei}}, \bibinfo {author} {\bibfnamefont {Y.}~\bibnamefont {Tang}}, \bibinfo
  {author} {\bibfnamefont {J.~L.}\ \bibnamefont {Grab}}, \bibinfo {author}
  {\bibfnamefont {J.}~\bibnamefont {Schubert}}, \bibinfo {author}
  {\bibfnamefont {D.~C.}\ \bibnamefont {Ralph}},\ and\ \bibinfo {author}
  {\bibfnamefont {D.~G.}\ \bibnamefont {Schlom}},\ }\bibfield  {title}
  {\bibinfo {title} {Structural, magnetic, and transport properties of
  {Fe}$_{\textrm{1-\textit{x}}}${Rh}$_{\textrm{\textit{x}}}$/{MgO}(001) films
  grown by molecular-beam epitaxy},\ }\href {https://doi.org/10.1063/1.5048303}
  {\bibfield  {journal} {\bibinfo  {journal} {Appl. Phys. Lett.}\ }\textbf
  {\bibinfo {volume} {113}},\ \bibinfo {pages} {082403} (\bibinfo {year}
  {2018})}\BibitemShut {NoStop}%
\bibitem [{\citenamefont {Shirane}\ \emph
  {et~al.}(1963{\natexlab{b}})\citenamefont {Shirane}, \citenamefont {Chen},
  \citenamefont {Flinn},\ and\ \citenamefont {Nathans}}]{1963Mssbauer}%
  \BibitemOpen
  \bibfield  {author} {\bibinfo {author} {\bibfnamefont {G.}~\bibnamefont
  {Shirane}}, \bibinfo {author} {\bibfnamefont {C.~W.}\ \bibnamefont {Chen}},
  \bibinfo {author} {\bibfnamefont {P.~A.}\ \bibnamefont {Flinn}},\ and\
  \bibinfo {author} {\bibfnamefont {R.}~\bibnamefont {Nathans}},\ }\bibfield
  {title} {\bibinfo {title} {M\"{o}ssbauer {Study} of {Hyperfine} {Fields} and
  {Isomer} {Shifts} in the {Fe}-{Rh} {Alloys}},\ }\href
  {https://doi.org/10.1103/PhysRev.131.183} {\bibfield  {journal} {\bibinfo
  {journal} {Phys. Rev.}\ }\textbf {\bibinfo {volume} {131}},\ \bibinfo {pages}
  {183} (\bibinfo {year} {1963}{\natexlab{b}})}\BibitemShut {NoStop}%
\bibitem [{\citenamefont {Xie}\ \emph {et~al.}(2020)\citenamefont {Xie},
  \citenamefont {Zhan}, \citenamefont {Hu}, \citenamefont {Hu}, \citenamefont
  {Chi}, \citenamefont {Zhang}, \citenamefont {Yang}, \citenamefont {Xie},
  \citenamefont {Zhu}, \citenamefont {Gao}, \citenamefont {Cheng},
  \citenamefont {Jiang},\ and\ \citenamefont {Li}}]{xie_2020}%
  \BibitemOpen
  \bibfield  {author} {\bibinfo {author} {\bibfnamefont {Y.}~\bibnamefont
  {Xie}}, \bibinfo {author} {\bibfnamefont {Q.}~\bibnamefont {Zhan}}, \bibinfo
  {author} {\bibfnamefont {Y.}~\bibnamefont {Hu}}, \bibinfo {author}
  {\bibfnamefont {X.}~\bibnamefont {Hu}}, \bibinfo {author} {\bibfnamefont
  {X.}~\bibnamefont {Chi}}, \bibinfo {author} {\bibfnamefont {C.}~\bibnamefont
  {Zhang}}, \bibinfo {author} {\bibfnamefont {H.}~\bibnamefont {Yang}},
  \bibinfo {author} {\bibfnamefont {W.}~\bibnamefont {Xie}}, \bibinfo {author}
  {\bibfnamefont {X.}~\bibnamefont {Zhu}}, \bibinfo {author} {\bibfnamefont
  {J.}~\bibnamefont {Gao}}, \bibinfo {author} {\bibfnamefont {W.}~\bibnamefont
  {Cheng}}, \bibinfo {author} {\bibfnamefont {D.}~\bibnamefont {Jiang}},\ and\
  \bibinfo {author} {\bibfnamefont {R.-W.}\ \bibnamefont {Li}},\ }\bibfield
  {title} {\bibinfo {title} {Magnetocrystalline anisotropy imprinting of an
  antiferromagnet on an amorphous ferromagnet in {FeRh}/{CoFeB}
  heterostructures},\ }\href {https://doi.org/10.1038/s41427-020-00248-x}
  {\bibfield  {journal} {\bibinfo  {journal} {NPG Asia Mater.}\ }\textbf
  {\bibinfo {volume} {12}},\ \bibinfo {pages} {67} (\bibinfo {year}
  {2020})}\BibitemShut {NoStop}%
\bibitem [{\citenamefont {Bergqvist}\ and\ \citenamefont
  {Bergman}(2018)}]{Bergqvist_2018}%
  \BibitemOpen
  \bibfield  {author} {\bibinfo {author} {\bibfnamefont {L.}~\bibnamefont
  {Bergqvist}}\ and\ \bibinfo {author} {\bibfnamefont {A.}~\bibnamefont
  {Bergman}},\ }\bibfield  {title} {\bibinfo {title} {Realistic finite
  temperature simulations of magnetic systems using quantum statistics},\
  }\href {https://doi.org/10.1103/PhysRevMaterials.2.013802} {\bibfield
  {journal} {\bibinfo  {journal} {Phys. Rev. Mater.}\ }\textbf {\bibinfo
  {volume} {2}},\ \bibinfo {pages} {013802} (\bibinfo {year}
  {2018})}\BibitemShut {NoStop}%
\bibitem [{\citenamefont {Cabassi}\ \emph {et~al.}(2010)\citenamefont
  {Cabassi}, \citenamefont {Bolzoni},\ and\ \citenamefont
  {Casoli}}]{cabassi_differential_2010}%
  \BibitemOpen
  \bibfield  {author} {\bibinfo {author} {\bibfnamefont {R.}~\bibnamefont
  {Cabassi}}, \bibinfo {author} {\bibfnamefont {F.}~\bibnamefont {Bolzoni}},\
  and\ \bibinfo {author} {\bibfnamefont {F.}~\bibnamefont {Casoli}},\
  }\bibfield  {title} {\bibinfo {title} {Differential method for sample holder
  background subtraction in superconducting quantum interference device
  ({SQUID}) magnetometry},\ }\href
  {https://doi.org/10.1088/0957-0233/21/3/035701} {\bibfield  {journal}
  {\bibinfo  {journal} {Meas. Sci. Technol.}\ }\textbf {\bibinfo {volume}
  {21}},\ \bibinfo {pages} {035701} (\bibinfo {year} {2010})}\BibitemShut
  {NoStop}%
\bibitem [{\citenamefont {Vries}\ \emph {et~al.}(2013)\citenamefont {Vries},
  \citenamefont {Loving}, \citenamefont {Mihai}, \citenamefont {Lewis},
  \citenamefont {Heiman},\ and\ \citenamefont {Marrows}}]{deVries_2013}%
  \BibitemOpen
  \bibfield  {author} {\bibinfo {author} {\bibfnamefont {M.~A.~d.}\
  \bibnamefont {Vries}}, \bibinfo {author} {\bibfnamefont {M.}~\bibnamefont
  {Loving}}, \bibinfo {author} {\bibfnamefont {A.~P.}\ \bibnamefont {Mihai}},
  \bibinfo {author} {\bibfnamefont {L.~H.}\ \bibnamefont {Lewis}}, \bibinfo
  {author} {\bibfnamefont {D.}~\bibnamefont {Heiman}},\ and\ \bibinfo {author}
  {\bibfnamefont {C.~H.}\ \bibnamefont {Marrows}},\ }\bibfield  {title}
  {\bibinfo {title} {Hall-effect characterization of the metamagnetic
  transition in {FeRh}},\ }\href
  {https://doi.org/10.1088/1367-2630/15/1/013008} {\bibfield  {journal}
  {\bibinfo  {journal} {New J. Phys.}\ }\textbf {\bibinfo {volume} {15}},\
  \bibinfo {pages} {013008} (\bibinfo {year} {2013})}\BibitemShut {NoStop}%
\bibitem [{\citenamefont {Fan}\ \emph {et~al.}(2010)\citenamefont {Fan},
  \citenamefont {Kinane}, \citenamefont {Charlton}, \citenamefont {Dorner},
  \citenamefont {Ali}, \citenamefont {de~Vries}, \citenamefont {Brydson},
  \citenamefont {Marrows}, \citenamefont {Hickey}, \citenamefont {Arena},
  \citenamefont {Tanner}, \citenamefont {Nisbet},\ and\ \citenamefont
  {Langridge}}]{Fan_2010}%
  \BibitemOpen
  \bibfield  {author} {\bibinfo {author} {\bibfnamefont {R.}~\bibnamefont
  {Fan}}, \bibinfo {author} {\bibfnamefont {C.~J.}\ \bibnamefont {Kinane}},
  \bibinfo {author} {\bibfnamefont {T.~R.}\ \bibnamefont {Charlton}}, \bibinfo
  {author} {\bibfnamefont {R.}~\bibnamefont {Dorner}}, \bibinfo {author}
  {\bibfnamefont {M.}~\bibnamefont {Ali}}, \bibinfo {author} {\bibfnamefont
  {M.~A.}\ \bibnamefont {de~Vries}}, \bibinfo {author} {\bibfnamefont
  {R.~M.~D.}\ \bibnamefont {Brydson}}, \bibinfo {author} {\bibfnamefont
  {C.~H.}\ \bibnamefont {Marrows}}, \bibinfo {author} {\bibfnamefont {B.~J.}\
  \bibnamefont {Hickey}}, \bibinfo {author} {\bibfnamefont {D.~A.}\
  \bibnamefont {Arena}}, \bibinfo {author} {\bibfnamefont {B.~K.}\ \bibnamefont
  {Tanner}}, \bibinfo {author} {\bibfnamefont {G.}~\bibnamefont {Nisbet}},\
  and\ \bibinfo {author} {\bibfnamefont {S.}~\bibnamefont {Langridge}},\
  }\bibfield  {title} {\bibinfo {title} {Ferromagnetism at the interfaces of
  antiferromagnetic {FeRh} epilayers},\ }\href
  {https://doi.org/10.1103/PhysRevB.82.184418} {\bibfield  {journal} {\bibinfo
  {journal} {Phys. Rev. B}\ }\textbf {\bibinfo {volume} {82}},\ \bibinfo
  {pages} {184418} (\bibinfo {year} {2010})}\BibitemShut {NoStop}%
\bibitem [{\citenamefont {Loving}\ \emph {et~al.}(2013)\citenamefont {Loving},
  \citenamefont {Jimenez-Villacorta}, \citenamefont {Kaeswurm}, \citenamefont
  {Arena}, \citenamefont {Marrows},\ and\ \citenamefont {Lewis}}]{Loving_2013}%
  \BibitemOpen
  \bibfield  {author} {\bibinfo {author} {\bibfnamefont {M.}~\bibnamefont
  {Loving}}, \bibinfo {author} {\bibfnamefont {F.}~\bibnamefont
  {Jimenez-Villacorta}}, \bibinfo {author} {\bibfnamefont {B.}~\bibnamefont
  {Kaeswurm}}, \bibinfo {author} {\bibfnamefont {D.~A.}\ \bibnamefont {Arena}},
  \bibinfo {author} {\bibfnamefont {C.~H.}\ \bibnamefont {Marrows}},\ and\
  \bibinfo {author} {\bibfnamefont {L.~H.}\ \bibnamefont {Lewis}},\ }\bibfield
  {title} {\bibinfo {title} {Structural evidence for stabilized ferromagnetism
  in epitaxial {FeRh} nanoislands},\ }\href
  {https://doi.org/10.1088/0022-3727/46/16/162002} {\bibfield  {journal}
  {\bibinfo  {journal} {J. Phys. D: Appl. Phys.}\ }\textbf {\bibinfo {volume}
  {46}},\ \bibinfo {pages} {162002} (\bibinfo {year} {2013})}\BibitemShut
  {NoStop}%
\bibitem [{\citenamefont {Suzuki}\ \emph {et~al.}(2011)\citenamefont {Suzuki},
  \citenamefont {Naito}, \citenamefont {Itoh}, \citenamefont {Sato},\ and\
  \citenamefont {Taniyama}}]{suzuki_2011}%
  \BibitemOpen
  \bibfield  {author} {\bibinfo {author} {\bibfnamefont {I.}~\bibnamefont
  {Suzuki}}, \bibinfo {author} {\bibfnamefont {T.}~\bibnamefont {Naito}},
  \bibinfo {author} {\bibfnamefont {M.}~\bibnamefont {Itoh}}, \bibinfo {author}
  {\bibfnamefont {T.}~\bibnamefont {Sato}},\ and\ \bibinfo {author}
  {\bibfnamefont {T.}~\bibnamefont {Taniyama}},\ }\bibfield  {title} {\bibinfo
  {title} {Clear correspondence between magnetoresistance and magnetization of
  epitaxially grown ordered {FeRh} thin films},\ }\href
  {https://doi.org/10.1063/1.3556754} {\bibfield  {journal} {\bibinfo
  {journal} {J. Appl. Phys.}\ }\textbf {\bibinfo {volume} {109}},\ \bibinfo
  {pages} {07C717} (\bibinfo {year} {2011})}\BibitemShut {NoStop}%
\bibitem [{\citenamefont {Raju}\ \emph {et~al.}(2019)\citenamefont {Raju},
  \citenamefont {Yagil}, \citenamefont {Soumyanarayanan}, \citenamefont {Tan},
  \citenamefont {Almoalem}, \citenamefont {Ma}, \citenamefont {Auslaender},\
  and\ \citenamefont {Panagopoulos}}]{raju_evolution_2019}%
  \BibitemOpen
  \bibfield  {author} {\bibinfo {author} {\bibfnamefont {M.}~\bibnamefont
  {Raju}}, \bibinfo {author} {\bibfnamefont {A.}~\bibnamefont {Yagil}},
  \bibinfo {author} {\bibfnamefont {A.}~\bibnamefont {Soumyanarayanan}},
  \bibinfo {author} {\bibfnamefont {A.~K.~C.}\ \bibnamefont {Tan}}, \bibinfo
  {author} {\bibfnamefont {A.}~\bibnamefont {Almoalem}}, \bibinfo {author}
  {\bibfnamefont {F.}~\bibnamefont {Ma}}, \bibinfo {author} {\bibfnamefont
  {O.~M.}\ \bibnamefont {Auslaender}},\ and\ \bibinfo {author} {\bibfnamefont
  {C.}~\bibnamefont {Panagopoulos}},\ }\bibfield  {title} {\bibinfo {title}
  {The evolution of skyrmions in {Ir}/{Fe}/{Co}/{Pt} multilayers and their
  topological {Hall} signature},\ }\href
  {https://doi.org/10.1038/s41467-018-08041-9} {\bibfield  {journal} {\bibinfo
  {journal} {Nat. Commun.}\ }\textbf {\bibinfo {volume} {10}},\ \bibinfo
  {pages} {696} (\bibinfo {year} {2019})}\BibitemShut {NoStop}%
\bibitem [{\citenamefont {Zadorozhnyi}\ and\ \citenamefont
  {Dahnovsky}(2023)}]{Zadorozhnyi_2023}%
  \BibitemOpen
  \bibfield  {author} {\bibinfo {author} {\bibfnamefont {A.}~\bibnamefont
  {Zadorozhnyi}}\ and\ \bibinfo {author} {\bibfnamefont {Y.}~\bibnamefont
  {Dahnovsky}},\ }\bibfield  {title} {\bibinfo {title} {Topological hall effect
  in three-dimensional centrosymmetric magnetic skyrmion crystals},\ }\href
  {https://doi.org/10.1103/PhysRevB.107.054436} {\bibfield  {journal} {\bibinfo
   {journal} {Phys. Rev. B}\ }\textbf {\bibinfo {volume} {107}},\ \bibinfo
  {pages} {054436} (\bibinfo {year} {2023})}\BibitemShut {NoStop}%
\bibitem [{\citenamefont {Karplus}\ and\ \citenamefont
  {Luttinger}(1954)}]{Karplus_1954}%
  \BibitemOpen
  \bibfield  {author} {\bibinfo {author} {\bibfnamefont {R.}~\bibnamefont
  {Karplus}}\ and\ \bibinfo {author} {\bibfnamefont {J.~M.}\ \bibnamefont
  {Luttinger}},\ }\bibfield  {title} {\bibinfo {title} {Hall effect in
  ferromagnetics},\ }\href {https://doi.org/10.1103/PhysRev.95.1154} {\bibfield
   {journal} {\bibinfo  {journal} {Phys. Rev.}\ }\textbf {\bibinfo {volume}
  {95}},\ \bibinfo {pages} {1154} (\bibinfo {year} {1954})}\BibitemShut
  {NoStop}%
\bibitem [{\citenamefont {Nagaosa}\ \emph {et~al.}(2010)\citenamefont
  {Nagaosa}, \citenamefont {Sinova}, \citenamefont {Onoda}, \citenamefont
  {MacDonald},\ and\ \citenamefont {Ong}}]{Nagaosa_2010}%
  \BibitemOpen
  \bibfield  {author} {\bibinfo {author} {\bibfnamefont {N.}~\bibnamefont
  {Nagaosa}}, \bibinfo {author} {\bibfnamefont {J.}~\bibnamefont {Sinova}},
  \bibinfo {author} {\bibfnamefont {S.}~\bibnamefont {Onoda}}, \bibinfo
  {author} {\bibfnamefont {A.~H.}\ \bibnamefont {MacDonald}},\ and\ \bibinfo
  {author} {\bibfnamefont {N.~P.}\ \bibnamefont {Ong}},\ }\bibfield  {title}
  {\bibinfo {title} {Anomalous hall effect},\ }\href
  {https://doi.org/10.1103/RevModPhys.82.1539} {\bibfield  {journal} {\bibinfo
  {journal} {Rev. Mod. Phys.}\ }\textbf {\bibinfo {volume} {82}},\ \bibinfo
  {pages} {1539} (\bibinfo {year} {2010})}\BibitemShut {NoStop}%
\end{thebibliography}%
\end{document}